\documentstyle[titlepage,twoside,12pt]{book}
\begin{document}

\begin{titlepage}

{\LARGE \bf Basics of Quantum Computation} \\

\hspace*{4cm} {\LARGE \bf (~Part 1~)} \\ \\ \\ \\
{\bf Elem\'{e}r E ~Rosinger} \\ \\
{\small \it Department of Mathematics \\ and Applied Mathematics} \\
{\small \it University of Pretoria} \\
{\small \it Pretoria} \\
{\small \it 0002 South Africa} \\
{\small \it eerosinger@hotmail.com}

\end{titlepage}

\setcounter{page}{0}

\newpage

\vspace*{10cm} ~~~~~~~~~~ Dedicated to Meda

\newpage

{\LARGE \bf Table of Contents} \\ \\ \\

{\large \bf Part 1} \\ \\

1~ What is the point in Quantum Computation \hfill 1

\medskip
~~~~1.1~ Preview \hfill 1

\medskip
~~~~1.2~ A First View of the Advantages \hfill 5

\medskip
~~~~1.3~ Is Physics Nothing Else But Computation ? \hfill 12

\medskip
2~ First Quantum Computations \hfill 15

\medskip
~~~~2.1~ Quantum Bits, or Qubits \hfill 15

\medskip
~~~~2.2~ Single Qubit Gates \hfill 18

\medskip
~~~~2.3~ Composite Quantum Systems and Entanglement \hfill 21

\medskip
~~~~2.4~ Multiple Qubit Gates \hfill 28

\medskip
~~~~2.5~ Classical Computations on Quantum Computers \hfill 30

\medskip
~~~~2.6~ Keeping Quantum Gates Simple \hfill 33

\medskip
3~ Two Strange Phenomena \hfill 39

\medskip
~~~~3.1~ No-Cloning \hfill 39

\medskip
~~~~3.2~ Teleportation \hfill 44

\medskip
4~ Bell's Inequalities \hfill 53

\medskip
~~~~4.1~ Boole Type Inequalities \hfill 56

\medskip
~~~~4.2~ The Bell Effect \hfill 58

\medskip
~~~~4.3~ Bell's Inequalities \hfill 61

\medskip
~~~~4.4~ Locality versus Nonlocality \hfill 64

\medskip
5~ The Deutsch-Jozsa Algorithm \hfill 67

\medskip
~~~~5.1~ A Simple Case of Quantum Parallelism \hfill 67

\medskip
~~~~5.2~ Massive Quantum Parallelism \hfill 69

\medskip
~~~~5.3~ The Deutsch Algorithm \hfill 72

\medskip
~~~~5.4~ The Deutsch-Jozsa Algorithm \hfill 75

\bigskip
Bibliography \hfill 79 \\ \\ \\

{\large \bf Part 2} \\ \\

\medskip
6~ Quantum Fourier Transform \hfill

\medskip
7~ The Grover Algorithm \hfill

\medskip
8~ The Shor Algorithm \hfill

\medskip
9~ Some Useful Properties \hfill

\medskip
10~ Conclusions \hfill

\medskip
Appendix 1~ Axioms of Quantum Mechanics \hfill

\medskip
~~~~A1.1~ State Space and Observables \hfill

\medskip
~~~~A1.2~ Six Axioms \hfill

\medskip
~~~~A1.3~ Types of Measurements \hfill

\medskip
~~~~A1.4~ Three Alternative Axioms \hfill

\medskip
~~~~A1.4~ Mathematical Failures of von Neumann's First Model \hfill

\medskip
Appendix 2~ Mathematical Tools \hfill

\medskip
~~~~A2.1~ The Dirac "bra-ket" Notation \hfill

\medskip
~~~~A2.2~ Eigenvalues and Eigenvectors \hfill

\medskip
~~~~A2.3~ Normal, Hermitian and Unitary Operators \hfill

\medskip
~~~~A2.4~ Spectral Representations \hfill

\medskip
~~~~A2.5~ Properties of Operators \hfill

\medskip
~~~~A2.6~ Tensor Products \hfill

\medskip
~~~~A2.7~ Abelian Groups, their Characters and Duals \hfill

\medskip
~~~~A2.8~ Finite Fourier Transforms and Complexity \hfill

\newpage

\pagenumbering{arabic}
\setcounter{page}{1}
\setcounter{chapter}{0}
\pagestyle{myheadings}
\markboth{E E Rosinger}{Basics of Quantum Computation}

\chapter{What is the point in Quantum Computation}

{\bf 1.1. Preview}

\bigskip
The literature on Quantum Computation has lately seen a number of textbooks published. Typically, they tend to be
rather encyclopedic, and thus quite lengthy as well, as they try to cover not only quantum computation proper, but also
a variety of related subjects, such as quantum decoherence, quantum error correction, quantum cryptography,
computational complexity, classical information theory, aspects of physical implementation, and so on. \\
Some of such books are simply collections of chapters written by various authors dealing with specific aspects of the
subject, and as such, serving not necessarily in the best way the unity and coherence of the presentation as a whole. \\

Such an involved approach, setting aside its merits, proves to have the obvious defect of making one's first time
access to the newly emerging realms of quantum computation so much more difficult. And this difficulty can be
experienced  even by a typical readership trained in science, such as mathematicians, physicists, or engineers, who
may wish to learn about the basics of quantum computation, and do so in a clear and rigorous enough manner, and not
merely on the level of science popularization. \\
Indeed, entering the subject of quantum computation does already present the usual science trained readership with {\it
three} quite inevitable major difficulties :  issues related to computational complexity, the strangeness of algorithms
for quantum computers, and above all, the strange and highly counter intuitive world of quantum phenomena in general. \\

The aim of this textbook is to {\it bridge} in regard of quantum computation what proves to be a considerable threshold
even to the usual science trained readership between the level of science popularization, and on the other hand, the
presently available more encyclopedic textbooks. \\
In this respect the present textbook is aimed to be a short, simple, rigorous and direct introduction, addressing itself
only to quantum computation proper. \\
There has been a certain tradition in the science literature in writing such introductions, albeit it may have been less
familiar lately. One of the examples which may come to one's mind is given by the well known Methuen monographs. \\

Quantum Computation presents the typical science trained reader with a {\it double novelty}, and also a {\it double
strangeness}. Namely, quantum physics is highly counter intuitive, and consequently, so are strikingly novel features
of the algorithms and the corresponding programs on quantum computers. \\
This textbook focuses as early as possible on the major new, typical, and so far exclusive {\it resources} of quantum
computers, given by such quantum phenomena as : \\

{\it superposition}, {\it entanglement}, {\it interference}, {\it parallelism}, and {\it reversible computation}. \\

A main issue, therefore, in quantum computation is that, as seen in Fig. 1.2.1 below, the algorithms and programs on
quantum computers only have a certain {\it limited} overlap with the usual algorithms and programs on electronic
digital computers. Indeed, on one hand, quite a number of usual algorithmic operations on electronic digital computers
are {\it not} available on quantum computers. On the other hand, quantum computers allow a number of algorithmic
operations which are incomparably more {\it powerful} than anything available on electronic digital computers. \\

The prerequisites in this textbook are those familiar for a large number of science trained readership. Namely, we
assume some basic knowledge about the way usual electronic digital computers process information represented by
classical binary bits. Also some familiarity is assumed with Linear Algebra, and in particular, with real or complex
vector spaces, their isomorphisms, linear mappings between such spaces, the representation of such mappings by
matrices, the eigenvectors and eigenvalues of such mappings or matrices, as well as the diagonalization of special
classes of such mappings or matrices. Certain minimal knowledge on tensor products of vector spaces, as well as on
finite Fourier transforms and complexity of computation will be required. However, all these subjects are reviewed for
the convenience of the reader in Appendix 2. \\
As in most of the literature on quantum physics and quantum computation, we shall use the so called "bra-ket" notation
of Dirac which proves to have important advantages. This notation is presented also in Appendix 2. \\
So much for the mathematical type prerequisites. \\

When it comes to physics, this is of course the main point in quantum computation, since whatever is new, and in fact,
quite spectacularly so in this respect, does come, and can only come, from those specific properties of quantum
systems which do not have any correspondent in classical physics, including usual electronics. \\
Here however, the situation is quite difficult as only a minority of the science trained readership is familiar with
quantum physics. And then, the approach in this textbook is to give in Appendix 1, six well known axioms of
quantum physics which will be sufficient for the presentation and understanding of the issues in quantum computation
dealt with in this textbook. Fortunately, these six axioms can be presented in terms of Linear Algebra, and do not need
additional detailed or involved physical arguments in order to be used in the rest of this textbook. \\

The two Appendices can be studied step by step, as the need arises during the reading of the main part of the book.
This is one reason why the material in them was not placed as an introductory part at the start of the textbook. \\

In this way, this textbook can be used starting with more advanced undergraduate students. However, the readership is
much wider, namely, all those trained in science who have some familiarity with usual electronic digital computers,
and may now wish to become familiar with quantum computation as well, without having to use as a first reading the
typical encyclopedic text available so far. \\

The content of this textbook is as follows. In the next section several of the more important novelties and advantages
of quantum computation are presented in short and in an informal manner. Chapter 2 introduces the very first specific
elements of quantum computation, namely, the so called qubits, quantum gates, and the all important phenomena of
superposition and entanglement. Immediately after, in chapter 3, two specific, rather strange and unexpected quantum
computation phenomena, namely, the so called no-cloning and teleportation are presented. Although these phenomena
appear to be quite different, their early introduction has the advantage to make the reader aware of some of the
important specifics of quantum computation. In chapter 4, the celebrated Bell inequalities are introduced. They play a
fundamental role in Quantum Mechanics, and as such cannot but have an important effect in quantum computation as
well. These chapters 2 - 4 form together the entrance to the subsequent presentation of specific algorithms typical for
quantum computation, algorithms which can be found in the following chapters 5 - 8. Such algorithm are indeed very
different from those we have been accustomed to when using usual electronic digital computers. Chapter 5 gives a
gradual insight into some of the applications of quantum parallelism and interference, starting with a simple case, and
ending with the full version of the Deutsch-Jozsa algorithm. Chapter 6 deals with the essentials of the theoretical
background of the Quantum Fourier Transform, which is then used in the Grover and Shor algorithms in chapters 7 and
8, respectively. The main part of textbook ends with several additional facts and comments in chapters 9 and 10. \\
As far as the two Appendices which complete the textbook, their content was mentioned earlier. \\ \\

{\bf 1.2 A First View of the Advantages}

\bigskip
Quantum computation has in certain impressive ways exploded upon us during the last decade. This comes more than
eight decades after the establishment by Max Planck in 1900 of Quantum Mechanics, the theory upon which quantum
computation is based. A number of initial insights, principles and results relevant for quantum computation were
obtained in the 1980s in works by R Feynman, D Deutsch and a few others, Brown, Deutsch [1-3], Feynman [1,2]. \\

A crucial moment of vast potential practical implications, however, occurred in 1994, when P Shor showed that quantum
computers can find the prime factors of large integers incomparably faster than usual electronic digital computers, thus
they may revolutionize the ways in which the coding of information is being done at present. This would of course lead
to a major challenge to the security of public-key crypto-systems upon which much of governmental and private
communication is based. \\
What prevents at present such a security challenge is the fact that, for the time being, there are not available large
enough quantum computers, that is, quantum mechanical devices which could effectively implement the massive
advantages already developed by the theory of quantum computation. \\

The Shor quantum algorithm for factorization in prime numbers, as mentioned later, is no less
than {\it exponentially} faster when compared with any other such algorithm known so far on
usual digital electronic computers. Another quite impressive breakthrough was Grover quantum
algorithm for search which is quadratically faster than any possible such algorithm on a usual
digital computer. \\
Such examples of highly practical interest have, no doubt, brought in a sharp focus the issue
of quantum computation, from the point of view of both theoretical and effective physical
implementation. \\

These massive advantages of quantum computation come precisely from the rather unusual, strange and surprising,
that is, far from classical properties of quantum mechanical systems. In particular, quantum mechanical systems can
behave in ways which are inconceivable in the case of electronic devices upon which the usual digital computers are
based. This fundamental difference between quantum mechanical devices, and on the other hand, all the other
classical ones, including electronic devices, is at the root of the massive power of quantum computing. \\

However, the comparative situation between classical and quantum computation is {\it not} quite that simple and
straightforward. Indeed, as mentioned in detail in the sequel, when going from usual electronic digital computers to
quantum computers, one not only {\it gains} a number of massive advantages, but also {\it loses} several particularly
useful and familiar classical ones. In this way in such a transition from usual to quantum computation, one enters
under the realm of the saying : \\

"You win some, you lose some ...",  \\

as illustrated in Fig. 1.2.1 below. However, as it turns out, what one loses is more than fully, and in fact, quite
spectacularly compensated by what one wins.

\newpage

\bigskip
\begin{math}
\setlength{\unitlength}{1cm}
\thicklines
\begin{picture}(15,10)
\put(2,4){\line(1,0){5}}
\put(2,7.5){\line(1,0){5}}
\put(2,4){\line(0,1){3.5}}
\put(7,4){\line(0,1){3.5}}
\put(3.5,6){$*1$}
\put(3,5){$\mbox{on~  bits}$}
\put(5.5,4.5){\line(1,0){5}}
\put(5.5,9){\line(1,0){5}}
\put(5.5,4.5){\line(0,1){4.5}}
\put(10.5,4.5){\line(0,1){4.5}}
\put(9,8){$*2$}
\put(9,7.5){$*3$}
\put(9,7){$*4$}
\put(9,6.5){$*5$}
\put(9,6){$*6$}
\put(8,5){$\mbox{on~  qubits}$}
\put(2,3.5){$\mbox{operations by}$}
\put(2,3.1){$\mbox{electronic computers}$}
\put(6.8,9.67){$\mbox{operations by}$}
\put(6.8,9.27){$\mbox{quantum computers}$}
\put(5,2){$\mbox{Fig. 1.2.1}$}
\put(3.5,1){$*1~~\mbox{irreversible~computation}$}
\put(3.5,0.5){$*2~~\mbox{superposition}$}
\put(3.5,0){$*3~~\mbox{entanglement}$}
\put(3.5,-0.5){$*4~~\mbox{interference}$}
\put(3.5,-1){$*5~~\mbox{parallelism}$}
\put(3.5,-1.5){$*6~~\mbox{reversible computation}$}
\end{picture}
\end{math} \\ \\ \\ \\

Let us start by noting that most of the operations performed by usual electronic digital computers are {\it irreversible}.
For instance, this holds for one of such basic operations like the addition of two integer numbers. Indeed, when we add
$a$ and $b$, and obtain $a + b$, we cannot in general recover from that sum the two initial terms $a$ and $b$. On the
other hand, as we shall see, the typical operations in quantum computers are given by {\it unitary} linear operators,
thus they are {\it reversible}. This follows from the axioms of Quantum Mechanics, according to which the dynamics of
a quantum system is always described by some unitary, thus invertible operator, unless some measurement is
performed. Of course, this does not mean that irreversible operations cannot at all be performed by quantum
computers. However, such operations are related to measurement processes in which the quantum system interacts
with a macroscopic measurement device. \\
Fortunately, this restriction on irreversible operations in the case of quantum computers can easily be compensated,
as will become obvious later. \\

Here it is important to note that, as seen in Appendix 1, according to the axioms of Quantum Mechanics a measurement
performed on a quantum mechanical system  does {\it not} always collapse the state of that system, does {\it not}
always have a probabilistic outcome, and is {\it not} always an irreversible process. However, typically, such a
measurement does collapse the state, its outcome is probabilistic, and it leads to an irreversible process. \\

As far as the new and unprecedented abilities quantum computers have owing to such typically quantum phenomena
like superposition, entanglement, interference and parallelism, we shall see the extent to which they revolutionize
computation by allowing a massive power. \\

Needless to say, the known laws of nature do not stop at those of electro-magnetism. And as it turns out, quantum
processes offer the possibility for a far more powerful computation. However, the classical laws of electro-magnetism,
on the one hand, and the laws of quantum processes, on the other hand, are vastly different, with the latter being also
highly surprising and counter intuitive, as they no longer relate to our every day experience. Consequently, when we
go from usual electronic digital computers to quantum computers, we have to develop completely {\it new approaches}
in computation. \\
This is actually what Quantum Computation is all about. \\

Related to the massive power, or speed of quantum computers, let us recall in somewhat more precise terms that from
the point of view of our usual electronic digital computers, problems get divided in two sharply different classes,
namely, of {\it polynomial}, respectively, {\it exponential complexity}, when it comes to the {\it number} of computer
operations involved in their solution. \\
A problem of polynomial complexity requires a computation time which in terms of the size, say $n$, of the respective
problem does not grow faster than a certain fixed power $n^k$ of that size, where $k$ is determined by the given
problem, but not by its size $n$. In particular, when $k = 1$, such problems are called of linear complexity. Such
problems, as well as more general ones of polynomial complexity for which $k$ is not too large, can easily be solved
on electronic digital computers even for considerable sizes for $n$. \\
Some typical examples are finding the smallest, or for that matter, the largest, number in a list of $n$ given numbers,
or performing the multiplication of two $n \times n$ matrices. For both of these problems $k \leq 2$. Another example is
the inversion of an $n \times n$ matrix which has nonzero determinant, for which $k \leq 3$. \\

On the other hand, a problem of exponential complexity requires a computation time which grows like an exponent
$k^n$ with the size $n$ of the respective problem. Here $k$ depends on the particular problem, but not on the size $n$
of that problem. And obviously, this leads to a tremendous growth even for $k = 2$, as the ancient story about the
origin of the chess game and of the corresponding remuneration problem of its inventor can attest. \\
Unfortunately, for a lot of important problems which one encounters in practical situations we could so far find only
algorithms of exponential complexity, and with $k \geq 2$. Among such problems are the so called {\it travelling
salesman's problem}, or the {\it factorization in prime numbers} of larger integers, with the second problem playing, as
mentioned, a fundamental role in present day coding, Brown. \\
By the way, recently, M Agrawal, N Kayal and N Saxena of the Indian Institute of Technology in Kanpur, claimed to have
a polynomial algorithm for testing whether a number is prime or not, Agrawal, et.al. \\

What P Shor managed to show in 1994 is that the factorization problem becomes of a mere polynomial, and in fact, of less
than cubic complexity, when solved with a quantum computer. More precisely, an n-bit number can be factorized in primes
with no more than in $O ( n^2 \log n \log \log n )$ steps, see Giorda, et.al., in Legget, et.al. \\
This is in sharp contradistinction with the algorithms known so far for this problem and aimed for usual electronic
digital computers. Indeed, such known algorithms do not have any comparably low complexity, not even even a polynomial one,
since the best so far among them, due to Pollard and Strassen, needs in general $O ( \exp ( C~ n^{1/3} \log^{2/3} n ) )$
steps, for a suitable constant $C > 0$. \\

And as the theory of quantum computation shows it in general, such an earlier hard to imagine massive reduction in the
complexity of problems, when one goes from usual electronic digital computers to quantum computers, can happen for rather
large classes of problems. \\

Needless to say, quantum computers prove to have a number of other important advantages as well, when compared
with the usual electronic digital ones. And from the point of view of a fuller understanding of such advantages we are
still in early stages of development, having so far found what may as well be but some of the first surprisingly powerful
possibilities and results. \\
In this respect it should not be overlooked that Quantum Mechanics itself, unlike the classical theories of physics upon
which the usual electronic digital computers are based, cannot be considered a closed theory, as among others, it is
still subject to fundamental controversies in the interpretation of its theoretical body. Consequently, it can be expected
that Quantum Mechanics may further witness important new developments which may as well impact upon quantum
computation. On the other hand, the recently emerged major interest in quantum computation, as
well as the developments related to its effective physical implementation which is still in early stages,
may bring in new points of view regarding the theory of Quantum Mechanics. This two way interaction can therefore be
expected to further contribute to the development of quantum computation. \\

One of the aims of this textbook is to make clear in sufficiently general, yet simple and direct terms such advantages of
quantum computation and quantum computers. However, this is not quite an immediate and trivial task because of the
following two reasons. \\
First, the dramatically increased computational powers of quantum computers come from specific, unique and
nonclassical, thus highly unusual and counter intuitive aspects of quantum mechanical systems, aspects upon which
such powers are essentially and directly based. It follows that one has to become familiar with some basics of
Quantum Mechanics in order to understand why, how and what quantum computers can, or for that matter, cannot do. In
Appendix 1 a short introduction to Quantum Mechanics is presented, which suffices for the purposes of this textbook.
Further reference is provided for those who may wish to go deeper in the related issues. \\
Second, and also as a consequence of the rather unusual ways of quantum systems, there are a number of operations
which quantum computers cannot do, although usual electronic digital ones can. This however is fully compensated by
what quantum computers can do, and they do so far beyond the abilities of usual electronic digital computers. In
particular, back in 1985, D Deutsch proved that quantum computers are {\it universal computers}, in other words, just
like the usual electronic digital computers, they can perform every algorithm, Brown. \\

The operations which quantum computers cannot do are again related to some of the unusual feature of quantum
systems. One of these features is that the time evolution of a quantum system is {\it reversible}, as long as no
measurement is performed on the system. On the other hand, a measurement of a quantum system will typically
{\it collapse} the state of that system, and do so in a probabilistic, rather than deterministic manner, leading to an
irreversible outcome. \\

The operations which quantum computers can do, and they can do them far beyond the performance of electronic digital
ones, come also from the unusual features of quantum systems, such as {\it entanglement}, {\it superposition}, {\it
parallelism}, or {\it interference}. \\
Further, one has to note that in the usual electronic digital computers the basic unit of information is the {\it bit}, which
can take two distinct values only. On the other hand, as we shall see in section 2.1, quantum systems allow for a far
{\it richer} basic unit of information which is called {\it quantum bit}, or for brevity, {\it qubit}. \\

In view of the above, it is clear that, when we want to solve a problem on a quantum computer, finding for it an
appropriate algorithm is not a trivial task, since we have to proceed in quite different ways than those which we use,
and by now are so much familiar with, in the case of an electronic digital computer. In this respect, for instance, the
algorithm of P Shor for prime number factorization gives a good example of the extent to which algorithms for quantum
computers may have to be rethought completely, and from their very start. \\

Finally, in addition to the mentioned conceptual nontriviality in using quantum computers, there is for the time being
also the practical limitation coming from the fact that the effective physical implementation of quantum computing has
not yet gone far enough, although progress in this regard is ongoing. \\
The challenge in building quantum computers, that is, actual physical systems which can perform quantum
computations, is that one may have to be able to control a certain suitable number, say, several hundred or perhaps
thousand, of individual quantum entities. This is of course far less easy than the classical electro-magnetic control of
flows of electrons through electric circuits in the microchips used in electronic digital computers. \\

The situation in Fig. 1.2.1 need of course {\it not} necessarily mean that we are now, or shall be in future, faced with an
either-or choice, namely, to use either usual electronic digital computers, or quantum computers. Indeed, it may prove
to be possible and convenient to use both of them, for instance, in a sort of {\it hybrid} setup, in which one can have
access to the comparative advantages of each of them. And for problems which do not present exponential complexity,
usual electronic digital computers can perform quite well, not to mention that there are plenty of well tested
corresponding algorithms and programs. Also, the writing of new such algorithms may be more easy, due to the
familiarity we have acquired over more than half a century, as well as to the fact that they need not be restricted mostly
to invertible gates, as it happens in the case of quantum computers. \\ \\

{\bf 1.3 Is Physics Nothing Else But Computation ?}

\bigskip
When it comes to effective means for implementing computation, and doing so outside of our human minds, we have so far
been obliged to make recourse to physical devices. In this regard, we can note three successive waves. The first was of
course {\it mechanical}, and it has ranged starting, for instance, from counting with small pieces of stone, from where
the very term Calculus happens to originate. It evolved to the more organized collections of such pieces which make up
an abacus, then in the 17th century it reached the mechanical sophistication of the machine constructed by the famous
French mathematician Blaise Pascal. Later, in the 19th century it even managed to overreach itself in the immense and
never completed Difference Engine of the English amateur scientist Charles Babbage. \\
The next and second wave starting in the 1940s, and represented by our present day usual {\it electronic} digital
computers, has been incomparably more advanced, and we are mostly still there, as the forthcoming third wave, of
{\it quantum} computers, is not yet at the stage where it could compete in practice. \\

Needless to say, these three successive waves have had a far wider impact upon human thinking and vision than in the
realms of computation only. After all, we have, especially after Newton, gone through a so called mechanical view of the
universe, and lately, since the 1920s, we tend to believe that everything is but a ... quantum cloud ... \\

As far as computation is concerned, its essential reliance in our times on the latest of the most basic laws of physics
has led to the question of the possible {\it identity} between physics and computation. More precisely, the question
emerged whether physics is, after all, nothing but an {\it information processing} done by Nature. And then, as Quantum
Mechanics happens to be the latest and most subtle of our theories of physics, the question arises whether or not the
Universe as a whole is but a quantum computer, Brown, Deutsch [1-3]. \\

One of perhaps the first such attempts to enquire into the possible identity between physics and computation was the
paper Simulating Physics With Computers, by the American physicist Richard Feynman, a famous Nobel Prize scientist. That
paper was delivered in 1981 at MIT, at the first ever major conference on physics and computation, Feynman [1,2]. \\
The point, as stressed by Feynman in that talk, was {\it not} merely to approximate fundamental physical processes on a
computer, but to see whether one can perform on a computer the very same information processing which goes on within
the respective physical processes, as they take place out there in Nature. \\

Specifically, Feynman asked whether our usual electronic digital computers can possibly do the information processing
which is involved in quantum phenomena. And based on a number of arguments following from the laws of Quantum
Mechanics, Feynman concluded that the information processing which typically goes on in quantum phenomena is so immense
that our usual electronic digital computer are {\it nowhere} near to be able to do the same. \\

In this way, the mentioned 1981 paper of R Feynman can be seen as the first major message on the dramatic relative
limitation of usual electronic digital computers, when compared to the potentialities of quantum computers, and thus,
of quantum computation. \\

Quantum computation is in this regard the development of the massive potentialities of quantum computers, when
compared with the capabilites of the usual electronic digital ones, potentialities highlighted among others by R
Feynman. \\ \\

\chapter{First Quantum Computations}

{\bf 2.1 Quantum Bits, or Qubits}

\bigskip
Information is based on difference, distinction, or discrimination. In its {\it classical} form, its basic unit
corresponding to its simplest possible form is one {\it bit}. This corresponds to a discrimination between two states
only, say, $0$ and $1$. For instance, one bit of information corresponds to knowing the state of an electronic device
which, by assumption, can only have one of {\it two} possible states. This means that we can write

\bigskip
(2.1.1) \quad $ one~ bit \in \{~ 0,~ 1 ~\} $

\medskip
and it is precisely such bits which are all that is processed by usual electronic digital computers. \\

On the other hand, the {\it qubit}, which is the basic unit of information processed by quantum computers, corresponds
to the states $|~ \psi >~$ of a quantum entity whose state space is a {\it complex two dimensional vector space}, that is,
$|~ \psi > ~\in {\bf C}^2$. Thus a qubit is given by the following {\it infinite} amount of classical information

\bigskip
(2.1.2) \quad $ one~ qubit ~=~ |~ \psi > ~=~ \alpha~ |~ 0 > ~+~  \beta~ |~ 1 > ~\in {\bf C}^2 $

\medskip
where $|~ 0 >,~ |~1 > ~\in {\bf C}^2$ denote an orthonormal basis in ${\bf C}^2$, and the complex numbers $\alpha,~
\beta \in {\bf C}$ satisfy the relation

\bigskip
(2.1.3) \quad $ |~ \alpha ~|^2 ~+~ |~ \beta~|^2 ~=~ 1 $

\medskip
However, since the states $|~ \psi >$ and $ e^{i \eta}|~ \psi >$, for all $\eta \in [0, 2 \pi]$, are equivalent from
quantum mechanical point of view, see Appendix 1, it follows that $\alpha,~ \beta$ in (2.1.2) have together only {\it two}
degrees of freedom, thus for instance, we can take them as

\bigskip
(2.1.4) \quad $ \alpha ~=~ \cos \theta,~~  \beta ~=~ e^{i \eta} \sin \theta,~~~ \eta,~ \theta \in [0, 2 \pi] $

\medskip
In this way, by comparing the classical bit in (2.1.1) with the qubit in  (2.1.2) - (2.1.4), we can note from the start the
considerably more rich, and in fact {\it doubly infinite} classical information content in one qubit, relative to the minimal
nontrivial finite information content in one bit. \\

Here, one of the strange quantum phenomena already shows up, namely, a phenomenon which is the subject of the
celebrated riddle of "Schr\"{o}-\\dinger's cat", Auletta. \\

Indeed, on the one hand, a quantum computer can effectively handle this doubly infinite information which is in a qubit,
this being done as the result of such typical quantum phenomena like {\it superposition}, {\it parallelism}, {\it
interference}, {\it entanglement} and so on. And such a handling of one qubit is as much the most simple and easy
basic operation in a quantum computer, as is the handling of a classical bit in a usual electronic digital computer. \\

Yet on the other hand, when it comes to {\it retrieve} as a {\it classical information} the information content in a qubit,
we have to effect what is called a {\it measurement} on the respective quantum system. And this will in general cause
the {\it collapse} of the respective wave function which gives the state $|~ \psi >$~ of the qubit in (2.1.2). Consequently,
the classical information which we shall be able to obtain will in general only be one single usual bit, for instance,
either knowing that the respective quantum system is in the state $|~ 0 >$, or on the contrary, that it is in the state
$|~ 1 >$, with each of the two states appearing with the respective {\it probabilities}

\bigskip
(2.1.5) \quad $ |~ \alpha ~|^2,~~~ |~ \beta ~|^2 $

\medskip
What is most important to note here is that, in spite of the appearance of the probabilities in (2.1.5), what we are
dealing here with in the case of the qubits in (2.1.2) - ( 2.1.5) is {\it not} at all a classical probabilistic system. Indeed, a
corresponding classical probabilistic system would have two states $A$ and $B$, and would manifest them with the
respective probabilities $p$ and $q$. However, that classical system would {\it always}, and most {\it certainly}, be in
one, and only in one, of the states $A$ or $B$. Thus the probabilistic aspect would only come from the fact that we do
not know in which of these two states the classical system happens to be, although that system is certainly always in
one and only one of its two states. A typical example of such a classical probabilistic system is the tossing of a coin. \\

On the other hand, in the case of a qubit as in (2.1.2) - (2.1.5), the respective quantum system is in general {\it not} in
any particular one of the states $|~ 0 >~$ or $|~ 1 >$. Instead, the quantum system is typically, and in a specific quantum
manner, in {\it both} of the states $|~ 0 >$ {\it and} $|~ 1 >$ at the same time, this being the meaning of {\it superposition}
of the respective two states in the case of a qubit. \\
The riddle of "Schr\"{o}dinger's cat" was invented by E Schr\"{o}dinger precisely in order to point out such strange,
highly counter intuitive and typically quantum phenomena. \\

Let us summarize the above two facts. A quantum computer can easily and simply handle qubits which can carry a
doubly infinite amount of classical information. When we retrieve classically such an information, we can only obtain
one single bit, and in general, we can do so only with a certain probability. \\
This is but one first typical example of "you win some, you lose some ..." illustrated in Fig. 1.2.1. However, as seen in
the sequel, it is already the source of a tremendous power of quantum computers, when compared with the usual
electronic digital ones. \\

Needless to say, already these two facts make it clear that setting up algorithms for quantum computers is highly
nontrivial, when compared with the customary ways of algorithms for usual electronic digital computers. \\

However, the difference, as seen later, between quantum computers and usual electronic digital ones is further
accentuated when it comes to handling an arbitrary finite number $n \geq 1$ of qubits. Indeed, specifying $n$ classical
bits amounts to giving one {\it single} integer $1 \leq m \leq 2^n$. On the other hand, owing to {\it quantum
superposition}, specifying $n$ qubits can lead to specifying {\it at the same time} and {\it simultaneously} no less than
$2^n$ integers, and in fact, much more, see (2.3.11), (2.3.12) in section 2.3. \\
This alone, therefore, can already give an idea about the surprising and significant increase in capabilities of quantum
computers. \\

Yet in order further to accentuate the fact that with quantum computers we are in a situation in which "we win some, and
we lose some", we also have to note the following. In a usual electronic digital computer we can read off the above
mentioned integer value $m$, and do so without having in any way whatsoever affected the $n$ classical bits which
define it uniquely. On the other hand, in a quantum computer, if we read off the contents of $n$ qubits which are in
superposition, we are inevitably coming under the axioms about quantum {\it measurement}, as already mentioned
above, and will therefore typically, even if not always, alter the respective multiple qubits by collapsing them, see the
end of section 2.3 for further details. \\

Fortunately, what "we win" with quantum computers will more than compensate for what "we lose" ... \\ \\

{\bf 2.2  Single Qubit Gates}

\bigskip
In analogy with usual electronic digital computer, we call a {\it gate} any quantum system which can process one or
more qubits. To be precise, such a quantum system will have as inputs and outputs states given by one or more qubits,
and it will process them according to the axioms of Quantum Mechanics. Therefore, outside of {\it measurements}, the
states of quantum systems are processed by unitary, thus invertible operators. It follows that quantum gates must have
the {\it same} number of input and output qubits. \\

\bigskip
We shall start with some of the simplest, yet useful quantum gates which process one input qubit into one output qubit.
The general form of such a quantum single qubit gate, say, $A$ is \\

\begin{math}
\setlength{\unitlength}{1cm}
\thicklines
\begin{picture}(15,3)
\put(2,1.4){$|~ \psi >$}
\put(3.5,1.5){\line(1,0){1.5}}
\put(5,0.5){\line(0,1){2}}
\put(5,0.5){\line(1,0){2}}
\put(5,2.5){\line(1,0){2}}
\put(7,0.5){\line(0,1){2}}
\put(7,1.5){\line(1,0){1.5}}
\put(9,1.4){$|~ \chi >$}
\put(5.8,1.4){$A$}
\put(5.1,-0.5){$\mbox{Fig. 2.2.1}$}
\end{picture}
\end{math} \\ \\

Here, as also always in the sequel, it is assumed that the information flows {\it from left to right} in quantum gates.
Therefore, there is no need for arrows to indicate the flow of information. Clearly, this simplification in the graphic
representation of quantum gates is made possible by the fact that quantum gates process qubits according to unitary,
and thus invertible operators. \\
In the case of the graphic representation of logical gates processing classical bits in electronic digital computers, it is
not convenient, and also often impossible, to make such an assumption on the flow of information. \\

Back to the quantum gate in Fig.2.2.1, we note that $|~ \psi >,~ |~ \chi > ~\in {\bf C}^2$ are qubits, while $A : {\bf C}^2
~\longrightarrow~ {\bf C}^2$ is a {\it unitary} linear operator, thus in particular, it is {\it invertible}. It is convenient to use
a matrix representation for the quantum gate $A$, namely

\bigskip
(2.2.1) \quad $ A ~=~ \left ( \begin{array}{l} a~~~~b \\ \\
                                                          c~~~~d
                                 \end{array} \right ) $

\medskip
in which case for qubits $~|~ \psi > ~=~ \alpha~ |~ 0 > ~+~ \beta~ |~ 1 >,~~ |~ \chi > ~=~ $ \\
$\gamma~ |~ 0 > ~+~ \delta~ |~ 1 >$, for which $A~|~ \psi > ~=~ |~ \chi >$, we shall have

\bigskip
(2.2.2) \quad $ \left ( \begin{array}{l} a~~~~b \\ \\
                                                          c~~~~d
                                 \end{array} \right )
                        \left ( \begin{array}{l} \alpha \\ \\
                                                           \beta
                                 \end{array} \right ) ~=~
                        \left ( \begin{array}{l} \gamma \\ \\
                                                           \delta
                                  \end{array} \right ) $

\medskip
The first example of quantum gate we consider is the quantum NOT gate, or in short, the q-NOT gate which is given by
the unitary matrix

\bigskip
(2.2.3) \quad $ X ~=~ \left ( \begin{array}{l} 0~~~~1 \\ \\
                                                                   1~~~~0
                                          \end{array} \right ) $

\medskip
It is easy to see that in view of (2.2.2), we obtain in this case

\bigskip
(2.2.4) \quad $ X~ ( \alpha~ |~ 0 > ~+~ \beta~ |~ 1 > ) ~=~ \beta~ |~ 0 > ~+~ \alpha~ |~ 1 > $

\medskip
in other words, the q-NOT gate simply switches between themselves the states $|~ 0 >$ and $|~ 1 >$. \\

Other useful quantum gates are given by the unitary matrices

\bigskip
(2.2.5) \quad $ Y ~=~ \left ( \begin{array}{l} 0~~~-i \\ \\
                                                                   i~~~~0
                                          \end{array} \right ) $

\medskip
and

\bigskip
(2.2.6) \quad $ Z ~=~ \left ( \begin{array}{l} 1~~~~~~0 \\ \\
                                                                   0~~~-1
                                          \end{array} \right ) $

\medskip
which act upon a given qubit according to

\bigskip
(2.2.7) \quad $ \begin{array}{l} Y~ ( \alpha~ |~ 0 > ~+~ \beta~ |~ 1 > ) ~=~ -i ( - \beta~ |~ 0 > ~+~ \alpha~ |~ 1 > ) \\ \\
                                                  Z~ ( \alpha~ |~ 0 > ~+~ \beta~ |~ 1 > ) ~=~ \alpha~ |~ 0 > ~-~ \beta~ |~ 1 >
                        \end{array} $

\medskip
The above $X,~ Y$ and $Z$ are called the Pauli matrices. Also, we shall encounter the Hadamard gate defined by the
unitary matrix

\bigskip
(2.2.8) \quad $ H ~=~ ( 1 / \sqrt 2 ) \left ( \begin{array}{l} 1~~~~~~1 \\ \\
                                                                                        1~~~-1
                                          \end{array} \right ) $

\medskip
Let us note that we have the following relations with respect to the repeated application of the above quantum gates

\bigskip
(2.2.9) \quad $ X^2 ~=~ Y^2 ~=~ Z^2 ~=~ H^2 ~=~ I $

\medskip
which means that each of the gates $X,~ Y,~  Z$ and $H$ are square roots of the identity matrix, and  corresponding
quantum gate $I$. \\

In general, in view of the fact that an arbitrary quantum gate $A$ in (2.2.1) is only subjected to the condition to be
unitary, it follows that there are {\it infinitely} many single qubit quantum gates. Indeed, the general form of a $2 \times
2$ unitary matrix, see Appendix 2, is given by

\bigskip
(2.2.10) \quad $ A ~=~ e^{i a}~ \left ( \begin{array}{l} \cos b~~~-i \sin b \\ \\
                                                                                 -i \sin b~~~\cos b
                                                        \end{array} \right )
                                               \left ( \begin{array}{l} \cos c~~~- \sin c \\ \\
                                                                                 \sin c~~~~~~\cos c
                                                         \end{array} \right )
                                                \left ( \begin{array}{l} e^{- i d}~~~ 0 \\ \\
                                                                                   0~~~~~~e^{i d}
                                                          \end{array} \right ) $

\medskip
where $a,~ b,~ c$ and $d$ are arbitrary real numbers. \\

This again is a considerable advantage over the situation with one bit input and one bit output logical gates in usual
electronic digital computers, where obviously, there are only {\it four} such gates $F : \{~ 0,~ 1 ~\} ~\longrightarrow~
\{~ 0,~ 1 ~\}$. And two of them are trivial, as they have the constant value $0$, respectively, $1$. The third is the identity
gate, while the fourth is the NOT gate which sends $0$ to $1$, and $1$ to $0$. \\

A further advantage of the representation in (2.2.10) is that it allows to {\it approximate} arbitrary one qubit quantum
gates $A$ by a fixed and finite number of such gates, corresponding to suitably chosen values of the parameters
$a,~ b,~ c$ and $d$. \\ \\

{\bf 2.3 Composite Quantum Systems and Entanglement}

\bigskip
Before considering multiple qubit gates in the next section, it is useful to have a look at the unusual manner quantum
systems become aggregated into composite ones. This feature is again unique to Quantum Mechanics and it leads to
one of the most powerful capabilities of quantum computers which is based on what is called {\it entanglement}, Auletta.
This term was initially suggested in the 1930s by E Schr\"{o}dinger in his comments to the celebrated 1935 paper of
Einstein-Podolski-Rosen, or in short, EPR. \\
In fact, the phenomenon of entanglement goes very deep into the nature of quantum processes, and it raises a whole
host of fundamental issues, among them that of {\it nonlocality}. The mentioned EPR paper was the first to bring
entanglement and its dramatic effects into focus, and it elicited a reaction which since then has seen more than one
million related published papers, Auletta. Of a major interest in this regard has been what is called "Bell's
inequalities", published in 1964, see Bell, Cushing \& McMullin, Maudlin. We shall consider in chapter 4 certain
aspects of this issue which are relevant to quantum computation. \\

Since we are dealing here with quantum computation, we can restrict ourselves to quantum systems which have as
states a finite number $n \geq 1$ of qubits, say

$$ |~ \psi_1 > ~=~ \alpha_1~ |~ 0 > ~+~ \beta_1~ |~ 1 >, .~.~.~ ,~ |~ \psi_n >~=~
                                                              \alpha_n~ |~ 0 > ~+~ \beta_n~ |~ 1 > ~\in {\bf C}^2 $$

\medskip
Thus the state spaces of such quantum systems are ${\bf C}^m$, for various finite and integer values of $m \geq 1$. \\
Here however, we have to be careful about how we find out the state space of $n$ qubits, that is, what is the value of
$m$ for the corresponding ${\bf C}^m$ in which the $n$ qubits range. Indeed, one of the surprising and significant
advantages of quantum computers already comes here to the fore, as mentioned at the end of section 2.1. \\

Given two quantum systems $S$ and $T$, with the respective state spaces ${\bf C}^n$ and ${\bf C}^m$, let us consider
them together, as forming a composite quantum system denoted by $S \otimes T$, even if they may on occasion be
functioning independently. \\
What is uniquely specific to Quantum Mechanics is that the state space of this composite quantum system $S \otimes
T$ will be given by the {\it tensor product}

\bigskip
(2.3.1) \quad $ {\bf C}^n \otimes {\bf C}^m $

\medskip
This is much unlike in Classical Mechanics, where the state space of a composite system is given by the Cartesian
product of their respective state spaces. \\

The effect of the tensor product in (2.3.1) is that the {\it dimension} of the state space of the composite quantum system
$S \otimes T$ is the {\it product} of the dimensions of their respective state spaces, since we have the isomorphism of
vector spaces

\bigskip
(2.3.2) \quad $ {\bf C}^n \otimes {\bf C}^m ~\simeq~ {\bf C}^{n m} $

\medskip
Here again for comparison, and in order to point out the difference, we can recall that in Classical Mechanics the
dimension of the state space of the composite of two system is the sum of the dimensions of their respective state
spaces, since as mentioned, the state space of this composite is given by the Cartesian product of the two state
spaces involved. In particular, for instance, if $S$ and $T$ were classical systems, then their classical composite
would have the state space ${\bf C}^n \times {\bf C}^m = {\bf C}^{n + m}$. And clearly $n m > n ~+~m$, starting with quite
small values of $n,~ m$, with the difference between $n m$ and $n ~+~ m$ increasing fast. \\

Returning to qubits, and with a view to multiple qubit gates, let us note the following consequence of (2.3.1), (2.3.2).
Suppose we are given $n$ quantum systems, each having its state described by the respective qubits $|~ \psi_1 >,
~ .~.~.~ , |~ \psi_n > ~\in {\bf C}^2$. Then the composite quantum system will have its states described by multiple qubits

\bigskip
(2.3.3) \quad $ |~ \psi > ~=~ ( |~ \psi_1 >,~ .~.~.~ , |~ \psi_n > )~\in {\bf C}^2 \otimes ~.~.~.~ \otimes {\bf C}^2 ~\simeq~
                                                                                                                                                                    {\bf C}^{2^n} $

\medskip
with the tensor product having $n$ factors, thus the dimension of the state space of the $n$ multiple qubits $|~ \psi >$
will be $2^n$. \\
On the other hand, in case we would have $n$ classical mechanical systems, each with the state space ${\bf C}^2$,
their composite would be ${\bf C}^2 \times ~.~.~.~ \times {\bf C}^2 ~\simeq~ {\bf C}^{2n}$, which is obviously much smaller,
as soon as $n \geq 3$. \\

In this way, the dimension of the state space of multiple qubits grows {\it exponentially}, as the power $2^n$, in the
number $n$ of qubits involved, while such a growth in dimension cannot be attained in such a simple manner in
Classical Mechanics. \\

For instance, if we consider $n$ classical bits $b_1, ~.~.~.~ , b_n \in \{~ 0, 1 ~\}$ then according to the Cartesian product
rule which operates in the classical context, we have $b = ( b_1, ~.~.~.~ , b_n ) \in \{~ 0, 1 ~\}^n$ for the corresponding
classical multiple bit. Therefore there are $2^n$ such distinct multiple classical bits. However, this does not compare in
any way with the {\it infinite} amount of multiple qubits $|~ \psi >~$ in (2.3.3) which can range over the whole of the $2^n$
complex dimensional vector space ${\bf C}^2 \otimes ~.~.~.~ \otimes {\bf C}^2 ~\simeq~ {\bf C}^{2^n}$, except for the vector
zero. And all these multiple qubits are distinct from quantum mechanical point of view, unless they are obtained from
one another by a transformation of the form $c~ |~ \psi >$, with $ c \in {\bf C},~ c \neq 0$. \\

To conclude for the moment, the state space $\{~ 0, 1 ~\}^n$ of $n$ classical bits is but a finite set which altogether has
{\it only} $2^n$ distinct elements. On the other hand, the state space ${\bf C}^2 \otimes ~.~.~.~ \otimes {\bf C}^2 ~\simeq~
{\bf C}^{2^n}$ of $n$ quantum qubits is a complex vector space, and as such, it has $2^n$ as its complex dimension.
Thus the state space ${\bf C}^{2^n}$ has {\it infinitely} many states which, according to the equivalence given by the
above transformation $|~ \psi > ~\longmapsto~ c~ |~ \psi >$, with $ c \in {\bf C},~ c \neq 0$, are all distinct from quantum
mechanical point of view. A more precise expression of this infinity is given at the end of this section. \\

With respect to (2.3.1) - (2.3.3) it is most important to note that it is precisely the presence of tensor products in the state
space of composite quantum systems, and the resulting multiplication of dimensions, which allow quantum computers
to accomplish the rather incredible feat in allowing algorithms which may abolish the difference between polynomial
and exponential complexity, a difference which although highly inconvenient, it is nevertheless unavoidable when
using electronic digital computers. The algorithm of P Shor, for instance, shows in the case of prime factorization that
one can turn a problem which, on usual electronic digital computers has so far only algorithms with a very high
complexity, into a problem of a significantly lower complexity, when solved on a quantum computer. \\

Finally, let us note that the reason why the state space of the composite of two quantum systems is given by a tensor,
rather than a Cartesian product is an immediate consequence of the {\it linearity} property of the states of quantum
systems, thus of their property to be able to have their states in {\it superposition}. Let us illustrate all that in the
simple case when we compose two quantum systems $S$ and $T$, each having its states given by a respective single
qubit. Of course, in this particular case the respective tensor product of the two state spaces has the same complex
dimension 4 as their Cartesian product has. Nevertheless, we analyze more closely this simple case in order to avoid
complications of a merely technical nature. Needless to say, in view of (2.3.2), as soon as at least one of the two state
spaces has complex dimension larger than 2, their respective tensor product will have a complex dimension larger
than that of their Cartesian product. \\

We start by noting that the quantum system $S$ can, among others, be in one of the single qubit states $|~ 0 >$ or
$|~ 1 >$. Similarly for the quantum system $T$. It follows that among the states of the composite quantum system $S
\otimes T$ are the double qubits

\bigskip
(2.3.4) \quad $ ( |~ 0 >, |~ 0 >),~~~ ( |~ 0 >, |~ 1 > ),~~~ ( |~ 1 >, |~ 0 > ),~~~ ( |~ 1 >, |~ 1> ) $

\medskip
And then the linearity property of the states, which holds for any quantum system, will immediately imply that $S
\otimes T$ must in addition also have as states all the possible {\it superpositions} given by the {\it linear}
combinations

\bigskip
(2.3.5) \quad $ \begin{array}{l} \alpha~ ( |~ 0 >, |~ 0 >) ~+~ \beta~ ( |~ 0 >, |~ 1 > ) ~+~ \gamma~ ( |~ 1 >, |~ 0 > ) ~+~  \\
                                                    ~+~ \delta~ ( |~ 1 >, |~ 1> )
                          \end{array} $

\medskip
with $\alpha,~ \beta,~ \gamma,~ \delta \in {\bf C}$, for which

\bigskip
(2.3.6) \quad $ |~ \alpha ~|^2 ~+~ |~ \beta ~|^2 ~+~ |~ \gamma ~|^2 ~+~ |~ \delta ~|^2 ~=~ 1 $

\medskip
Since in Quantum Mechanics any nonzero state $|~ \psi >$ is equivalent with any other state $c~ |~ \psi >$, with $c \in
{\bf C},~ c \neq 0$, we can consider (2.3.5) alone, without the normalizing condition (2.3.6). In this way, it follows that the
state space of the two qubit composite quantum system $S \otimes T$ is indeed ${\bf C}^2 \otimes {\bf C}^2$, as
specified in general in (2.3.1). \\

In order to clarify the phenomenon of entanglement, let us now return to the general case in (2.3.1) of the composite
$S \otimes T$ of two quantum systems $S$ and $T$. We can assume that the state space ${\bf C}^n$ of $S$ has an
orthonormal basis $|~ 1 >,~ .~.~.~ , |~ n >$, while the state space ${\bf C}^m$ of $T$ has an orthonormal basis $|~ 1 >,
~ .~.~.~ , |~ m >$. Then every state $|~ \psi >~$ of $S$ and $|~ \chi >~$ of $T$ can be written respectively as

\bigskip
(2.3.7) \quad $ \begin{array}{l} |~ \psi > ~=~ \alpha_1~ |~ 1 > ~+~ .~.~. ~+~ \alpha_n~ |~ n > \\ \\
                                                    |~ \chi > ~=~ \beta_1~ |~ 1 > ~+~ .~.~. ~+~ \beta_m~ |~ m >
                          \end{array} $

\medskip
with $\alpha_1, ~.~.~.~ ,  \alpha_n,~ \beta_1, ~.~.~.~ , \beta_m \in {\bf C}$. \\
On the other hand, every state $|~ \phi >~$ of the composite quantum system $S \otimes T$ can be written as

\bigskip
(2.3.8) \quad $ |~ \phi > ~=~ \gamma_1~ |~ 1 > ~\otimes~ |~ 1 > ~+~ ~.~.~. ~+~ \gamma_{n m}~ |~ n > ~\otimes~ |~ m > $

\medskip
with $\gamma_1, ~.~.~.~ , \gamma_{n m} \in {\bf C}$. \\

Here we used the customary notation according to which a double qubit $( |~ i > , |~ j > )$ is also written as $|~ i >
~\otimes~ |~ j >$, or $|~ i > |~ j >$, and even simply as $|~ i, j >$, or $|~ i j >$, when this does not create confusion. \\

And now an essential feature of tensor products comes into play. Namely, by far most of the states $|~ \phi >~$ of the
composite quantum system $S \otimes T$ are {\it not} of the simple and particular form

\bigskip
(2.3.9) \quad $ |~ \phi > ~=~ |~ \psi > ~\otimes~ |~ \chi > $

\medskip
where $|~ \psi >$ and $|~ \chi >$ are states of the component systems $S$ and $T$, respectively. For instance, in the
case of double qubits, it can be seen easily that in ${\bf C}^2 \otimes {\bf C}^2$ we have

\bigskip
(2.3.10) \quad $ \begin{array}{l}  |~ 0, 1 > ~+~ |~ 1, 0 > ~\neq~ \\ \\
                                                     ~\neq~ ( \alpha~ |~ 0 > ~+~ \beta~ |~ 1 > ) \otimes ( \gamma~ |~ 0 > ~+~ \delta~ |~ 1 > )
                          \end{array} $

\medskip
for any values of $\alpha,~ \beta,~ \gamma,~ \delta \in {\bf C}$. \\

The states $|~ \phi >$ of a composite quantum system $S \otimes T$ for which (2.3.9) does not hold are called {\it
entangled}. And as noted, such entangled states constitute by far the majority, or in other words, the typical states in a
composite quantum system. \\

One such example of entangled state in a composite quantum system is the double qubit $|~ 0, 1 > ~+~ |~ 1, 0 >~$ in
${\bf C}^2 \otimes {\bf C}^2$, see (2.3.10). \\

Returning to the $n$ qubit systems in (2.3.3) with their states $|~ \psi > ~\in {\bf C}^{2^n}$, let us note that we have their
representations

\bigskip
(2.3.11) \quad $ |~ \psi > ~=~ \Sigma_{x_1, ~.~.~.~ ,~ x_n}~ \alpha_{x_1, ~.~.~.~ ,~ x_n}~ |~ x_1, ~.~.~.~ ,~ x_n > ~\in {\bf C}^{2^n} $

\medskip
where the sum is taken over all $x_1, ~.~.~.~ ,~ x_n \in \{~ 0, 1 ~\}$, while $\alpha_{x_1, ~.~.~.~ ,~ x_n} \in {\bf C}$ are
subject to the condition

\bigskip
(2.3.12) \quad $  \Sigma_{x_1, ~.~.~.~ ,~ x_n}~ | \alpha_{x_1, ~.~.~.~ ,~ x_n} |^2 ~=~ 1 $

\medskip
And in order to obtain in (2.3.11) different quantum states $ |~ \psi >$, that is, different $n$ qubits, the respective sets of
$\alpha_{x_1, ~.~.~.~ ,~ x_n}$ in two qubits given by (2.3.11) have to differ more than merely by a factor
$c \in {\bf C}$, with $| c | = 1$. \\
Clearly, the multiple infinity of such $n$ qubits $|~ \psi >$ goes far beyond the finite number of $2^n$ classical bits
of a classical $n$ bit system. Indeed, $n$ classical bits can only have $2^n$ different states. On the other hand, $n$
qubits can range, within condition (2.3.12), over a $2^n$ complex dimensional complex vector space, and they will all
give different states, as long as they differ by more than a factor $c \in {\bf C}$, with $| c | = 1$. \\

Here again, let us note that a quantum system which handles $n$ entangled qubits does in effect process such a
multiple infinity information as contained in (2.3.11) under the above mentioned conditions. And the respective quantum
system, based on the laws of Quantum Mechanics, processes such an infinite information just as simply as the usual
electronic digital computers do with the classical information, based on the Classical Mechanics. \\
The problem arises, as with "Schr\"{o}dinger's cat", when we want to retrieve in a classical manner that infinite amount
of information contained in a quantum system. In such a case, as seen in section 2.1, we have to make a {\it quantum
measurement}, with all the consequent randomness and loss of information which in such a situation will happen
typically. \\

As far as quantum measurement is concerned in the context of multiple qubits in (2.3.11), (2.3.12), we can note the
following. According to the axioms of Quantum Mechanics, when such an $n$ qubit $|~ \psi >$ is measured, we shall
typically obtain {\it one} and only one set of $n$ classical bits $( x_1, ~.~.~.~ ,~ x_n ) \in \{~ 0, 1 ~\}^n$, and do so with the
respective probability $| \alpha_{x_1, ~.~.~.~ ,~ x_n} |^2$. \\
Furthermore, as an effect of measurement, the superposition of the large number of states in (2.3.11) will {\it collapse}
onto the corresponding state $|~ x_1, ~.~.~.~ ,~ x_n >$. Thus the ability of the quantum computer to handle {\it
simultaneously} all the states in the superposition in (2.3.11) will end. \\
Finally, due to the large number of terms in (2.3.12), it is often that such a probability $| \alpha_{x_1, ~.~.~.~ ,~ x_n} |^2$
is small. \\ \\

{\bf Practical Remark} \\

In view of the above, when setting up algorithms for quantum computers, it is useful to avoid an early loss of
superposition. This therefore means avoiding an early measurement. As far as enhancing the probability of the results
of measurements, this can be obtained by a judicious choice of quantum gates, that is, of unitary operators acting on
multiple qubits. \\
All this, however, need not mean that measurements have to be left up to the very end of such quantum algorithms.
Indeed, as seen for instance in the case of the algorithm for quantum teleportation in Fig. 3.2.1 in chapter 3, it can
happen that an appropriate measurement, leading as it always does to a classical information, can be useful not only
at the end of a quantum algorithm. \\ \\

{\bf 2.4 Multiple Qubit Gates}

\bigskip
Although there are an infinity of single qubit gates, there are obvious advantages in considering as well multiple qubit
gates. Here however we have to recall that in the case of quantum gates one has to have the same number of qubits
both at input and output. This is contrary to what happens with logical gates processing classical bits, used in
electronic digital computers, where for instance, the gates AND and OR each have two bits as input, and only one bit as
output. \\

A first quantum gate with two qubit input and two qubit output which we consider is the controlled-NOT, or simply CNOT
gate \\

\bigskip
\begin{math}
\setlength{\unitlength}{1cm}
\thicklines
\begin{picture}(15,3)
\put(2,2.8){$|~ \psi >$}
\put(3.5,2.9){\line(1,0){5}}
\put(9,2.8){$|~ \psi >$}
\put(2,0.1){$|~ \chi >$}
\put(3.5,0.2){\line(1,0){5}}
\put(9,0.1){$|~ \psi > ~\oplus~ |~ \chi >$}
\put(6,2.85){\circle*{0.5}}
\put(6,0.2){\line(0,1){2.6}}
\put(6,0.2){\circle{0.5}}
\put(5.1,-1){$\mbox{Fig. 2.4.1}$}
\end{picture}
\end{math} \\ \\

\medskip
which operates according to

\bigskip
(2.4.1) \quad $ \begin{array}{l} |~ 0~ 0 > ~\longmapsto~ |~ 0~ 0 >,~~~~~~ |~ 0~ 1 > ~\longmapsto~ |~ 0~ 1 > \\ \\
                                                    |~ 1~ 0 > ~\longmapsto~ |~ 1~ 1 >,~~~~~~ |~ 1~ 1 > ~\longmapsto~ |~ 1~ 0 >
                          \end{array} $

\medskip
thus when $|~ \psi > ~=~ |~ 0 >$, then $|~ \psi > ~\oplus~ |~ \chi > ~=~ |~ \chi >$, while for $|~ \psi > ~=~ |~ 1 >$, we obtain
$|~ \psi > ~\oplus~ |~ \chi > ~=~ X~ |~ \chi >$, see (2.2.4). The matrix representation of the operation of the CNOT gate is
therefore

\bigskip
(2.4.2) \quad $ \left ( \begin{array}{l} 1~~~ 0~~~ 0~~~ 0 \\
                                                             0~~~ 1~~~ 0~~~ 0 \\
                                                             0~~~ 0~~~ 0~~~ 1 \\
                                                             0~~~ 0~~~ 1~~~ 0
                                    \end{array} \right )
                            \left ( \begin{array}{l} \alpha \\
                                                               \beta \\
                                                               \gamma \\
                                                               \delta
                                      \end{array} \right )
                            ~=~
                            \left ( \begin{array}{l} \alpha \\
                                                               \beta \\
                                                               \delta \\
                                                               \gamma
                                      \end{array} \right ) $

\medskip
assuming that $|~ \psi > ~=~ \alpha~ |~ 0 > ~+~ \beta~ |~ 1 >,~ |~ \chi > ~=~ \gamma~ |~ 0 > ~+~ \delta~ |~ 1 >$. It is easy to
check that the above matrix is indeed unitary. \\ \\

{\bf Remark} \\

The special importance of the CNOT gate comes from the fact that any multiple qubit gate can be obtained as a
composition of CNOT gates and single qubit gates, see section 2.6. \\

This result about quantum gates corresponds to the classical result according to which every logical gate operating on
bits can be obtained from the composition of NAND gates. \\
Here we recall that a NAND gate operates on two classical bits $a,~ b$ according to $~~a ~NAND~ b ~=~ NOT~ ( a ~AND~
b )$. \\ \\

{\bf 2.5 Classical Computations on Quantum Computers}

\bigskip
As we mentioned, D Deutsch showed in 1985 that quantum computation, just like the usual electronic digital one, is
{\it universal}. Here we shall address in short some of the related issues. Namely, as we have seen, quantum gates
operate on qubits in a {\it reversible} manner, while classical logical gates operate on bits, and do so most often in an
{\it irreversible} way. \\

Therefore the question arises how can quantum gates process information in equivalent ways with classical logical
gates ? In other words, how can one turn irreversible operations into reversible ones ? \\

At a first thought, and on a rather metaphysical level, one could expect that quantum computers can indeed perform
classical computations. After all, it is a fundamental thesis of modern Physics that quantum phenomena underlie the
macroscopic ones, thus including the classical logical gates of usual electronic digital computers. However, since here
we are not dealing with metaphysics, we shall instead give a precise indication about the way classical computations
can be performed on quantum computers. \\

As it happens, the idea of a reversible computation appeared as a consequence of studying the problem of the
minimum energy needed in computation on usual electronic digital computers, Brown. One of the first steps in
clarifying this minimum energy was taken in 1949 by John von Neumann. \\
In 1961, Rolf Landauer made a crucial discovery by showing that the only processes in a computation which are
irreversible are those which erase information. This was to lead to the idea of reversible computation even before the
emergence of quantum conputation. Results in this respect were obtained by Yves Lecerf in 1963, and in their complete
form by Charles Bennett in 1973. Not much later, Ed Fredkin and Tom Toffoli showed independently the way to build
reversible computers. \\
It is however important to note that, by avoiding to erase information one creates, and also must carry along a
significant, if not even growing amount of redundancy, this being one of the prices one has to pay for reversible
computation. \\
Needles to say that at the time, such studies concerned not the quantum, but only the classical forms of computation,
that is, by electronic digital computers. \\
Further details regarding reversible computation can be found in Brown, Deutsch [1-3], Hirvensalo, Alber et.al. \\

The relevant result with respect to the questions formulated above is that the information processing by any classical
logical gate can be reproduced with the use of Toffoli gates which are reversible. \\

The Toffoli gate has three bits as input, and also three bits as output, namely, for classical bits $a, b, c \in \{~ 0, 1 ~\}$,
we have  \\

\bigskip
\begin{math}
\setlength{\unitlength}{1cm}
\thicklines
\begin{picture}(15,5)
\put(2.8,4.8){$a$}
\put(3.5,4.9){\line(1,0){5}}
\put(9,4.8){$a$}
\put(2.8,2.35){$b$}
\put(3.5,2.5){\line(1,0){5}}
\put(9,2.35){$b$}
\put(6,4.85){\circle*{0.5}}
\put(6,2.5){\line(0,1){2.4}}
\put(6,2.5){\circle*{0.5}}
\put(2.8,0){$c$}
\put(3.5,0.1){\line(1,0){5}}
\put(9,0){$c ~\oplus~ a b $}
\put(6,0.11){\circle{0.5}}
\put(6,0.1){\line(0,1){2.4}}
\put(5.1,-1.3){$\mbox{Fig. 2.5.1}$}
\end{picture}
\end{math} \\ \\ \\

\bigskip
where in the term $c \oplus a b$, the operation $\oplus$ is addition modulo $2$, while $a b$ is the usual multiplication.
In this way, written as an input-output table, the Toffoli gate has the form

\bigskip
(2.5.1) \quad $ \begin{array}{l} ( 0, 0, 0 ) ~\longmapsto~ ( 0, 0, 0 ) \\
                                                    ( 0, 0, 1 ) ~\longmapsto~ ( 0, 0, 1 ) \\
                                                    ( 0, 1, 0 ) ~\longmapsto~ ( 0, 1, 0 ) \\
                                                    ( 0, 1, 1 ) ~\longmapsto~ ( 0, 1, 1 ) \\
                                                    ( 1, 0, 0 ) ~\longmapsto~ ( 1, 0, 0 ) \\
                                                    ( 1, 0, 1 ) ~\longmapsto~ ( 1, 0, 1 ) \\
                                                    ( 1, 1, 0 ) ~\longmapsto~ ( 1, 1, 1 ) \\
                                                    ( 1, 1, 1 ) ~\longmapsto~ ( 1, 1, 0 )
                          \end{array} $

\medskip
It is easy to see that applying twice the Toffoli gate gives the identity. Thus the Toffoli gate is invertible, being its own
inverse. Consequently, the operation of the Toffoli gate is indeed reversible. \\

It is important to note that the {\it redundancy} in the output of the Toffoli gate which reproduces identically the bits $a$
and $b$ is the way to avoid {\it erasing} information, which according to Landauer, is a necessary condition for allowing
for {\it reversibility}. \\

In order to prove that every classical logical gate can be obtained from Toffoli gates it suffices to show that the NAND
gate can be constructed in that way. Indeed, we have \\

\bigskip
\begin{math}
\setlength{\unitlength}{1cm}
\thicklines
\begin{picture}(15,5)
\put(2.8,4.8){$a$}
\put(3.5,4.9){\line(1,0){5}}
\put(9,4.8){$a$}
\put(2.8,2.35){$b$}
\put(3.5,2.5){\line(1,0){5}}
\put(9,2.35){$b$}
\put(6,4.85){\circle*{0.5}}
\put(6,2.5){\line(0,1){2.4}}
\put(6,2.5){\circle*{0.5}}
\put(2.8,0){$1$}
\put(3.5,0.1){\line(1,0){5}}
\put(9,0){$1 ~\oplus~ a b ~=~ a~ \mbox{NAND}~ b$}
\put(6,0.11){\circle{0.5}}
\put(6,0.1){\line(0,1){2.4}}
\put(5.1,-1.3){$\mbox{Fig. 2.5.2}$}
\end{picture}
\end{math} \\ \\ \\

\bigskip
The classical Toffoli gate in Fig. 2.5.1 or (2.5.1) has a quantum gate version as well. Indeed, each triplet of classical
bits $(a, b, c ) \in \{~ 0, 1 ~\}^3$ can be uniquely associated with the quantum triplet $|~ a, b, c > ~\in {\bf C}^2 \otimes
{\bf C}^2 \otimes {\bf C}^2 \simeq {\bf C}^8$. And then (2.5.1) defines a unique $8 \times 8$ unitary matrix together with
the corresponding unitary operator $T : {\bf C}^8 \longrightarrow {\bf C}^8$ which gives the quantum Toffoli gate. And the
operations of this quantum Toffoli gate clearly contain as a particular case those of the classical Toffoli gate. \\

Finally, let us note that quantum computation can also simulate nondeterministic classical computation. For that
purpose, as is known, it is sufficient to simulate the randomness of a fair coin toss. This however can be done trivially,
by sending the quantum state $|~ 0 >$ through a Hadamard gate $H$, see (2.2.8). Indeed, we shall have then $H~ |~ 0 >
~=~ (1 / \sqrt ( 2 ) ) (~ |~ 0 > ~+~ |~ 1 > ~)$, thus by measuring this resulting state we shall obtain $|~ 0 >~$ or $|~ 1 >$, each
with probability $1/2$. \\ \\

{\bf2.6~ Keeping Quantum Gates Simple}

\bigskip
Let us recapitulate. \\

On usual electronic digital computers the smallest amount of information, as seen in (2.1.1), is one classical {\it bit}
which can be represented as an element of the {\it two} element set $\{~ 0, 1 ~\}$. It follows that in such a computer any
classical {\it logical} gate operating on one classical bit is given by one of the {\it four} functions $f : \{~ 0, 1 ~\}
\longrightarrow \{~ 0, 1 ~\}$. \\

On the other hand, in quantum computers, the smallest amount of information is a {\it qubit}, see (2.1.2) - (2.1.4)

\bigskip
(2.6.1) \quad $ |~ \psi > ~=~ \cos \theta~ |~ 0 > ~+~ e^{i \eta} \sin \theta~ |~ 1 > ~\in {\bf C}^2,~~~ \eta,~ \theta \in [ 0, 2 \pi ] $

\medskip
of which there are therefore a {\it double infinity}. \\
Now the {\it quantum} gates which operate on such single qubits are given by {\it unitary} operators, see (2.2.1)

\bigskip
(2.6.2) \quad $ A : {\bf C}^2 ~\longrightarrow~ {\bf C}^2 $

\medskip
of which there are a {\it quadruple infinity}, as follows form (2.2.10). \\
Let us recall that, therefore, each of such one qubit quantum gates $A$, which in the case of quantum computers are the
{\it simplest} possible gates, already processes at each step a double infinity of information, as given in (2.6.1). Such a
performance is of course impossible on usual electronic digital computers, where there cannot be any logical gates
which could in one single step process an infinite amount of information. \\
On the other hand, when extracting classical information from a quantum computer, and in particular, when we do so
from any given qubit (2.6.1), we can only obtain one classical bit, namely, one of the states $|~ 0 >$ or $|~ 1 >$. This
follows from the axioms of Quantum Mechanics relating to measurement. \\

Now, as seen already in section 2.4, and in more detail later, quantum algorithms may need quantum gates which
operate on multiple qubits as well. And as follows from (2.3.3), and the axioms of Quantum Mechanics, a quantum gate
which operates on $n$ qubits is given by an arbitrary {\it unitary} operator

\bigskip
(2.6.3) \quad $ U : {\bf C}^{2^n} ~\longrightarrow~ {\bf C}^{2^n} $

\medskip
Related to this, let us recall the uniquely convenient feature of quantum computers seen in (2.3.11). According to that, if
we take $n$ single qubits, each of them having only two states $|~ 0 >$ and $|~ 1 >$, and construct from them one
$n$-qubit composite system, then this system will have no less than $2^n$ different and {\it linearly independent}
states $|~ x_1, ~.~.~.~ , x_n >$, with $x_1, ~.~.~.~ , x_n \in \{~ 0, 1 ~\}$, which form the basis of the correpsonding $2^n$
dimensional complex vector space ${\bf C}^{2^n} $. And an $n$-qubit quantum gate $U$ in (2.6.3) can in general operate
{\it simultaneously} on all of these $2^n$ different and linearly independent states. \\ \\

{\bf Remark} \\

In this way, quantum gates on multiple qubits present {\it two} major advantages over usual logical gates. First, they
can operate on an {\it infinite} amount of information, and second, the number of quantum states on which they can
operate simultaneously grows {\it exponentially}, namely, like $2^n$, with the length $n$ of the number of qubits they
operate. \\

Again, however, and due to the same axioms of Quantum Mechanics, when we measure the effect of such a quantum
gate $U$, we shall only obtain $n$ classical bits, namely, one specific single state $|~ x_1, ~.~.~.~ , x_n >$. \\

Obviously, the infinite multiplicity of all such possible quantum gates in (2.6.3) is fast growing with $n$. Thus the {\it
practical} problem arises whether such $n$-qubit gates can be modelled, or at least approximated, by a small number
of quantum gates, each operating only on a small number of qubits.

\hfill $\Box$

\bigskip
Fortunately, we have a number of strong results in this respect and we shall recall several of them here. Further details
can be found in Alber et.al., Pittenger, and the literature cited there. \\
The general intuitive idea underlying such results is that unitary operators are in certain sense generalized {\it
rotations}. And as such, they should be reproducible in suitable ways by a composition of the {\it simplest} rotations,
which therefore are only supposed to involve {\it two} dimensions. \\

A result already mentioned section 2.4, is the following. Arbitrary $n$-qubit quantum gates $U$ in (2.6.3) can be {\it
constructed} form CNOT gates operating on two qubits, see Fig. 2.4.1, and the simplest quantum gates $A$ in (2.6.2)
which operate on a single qubit. \\

The precise details are as follows. Let us take any $n \geq 1$ fixed. \\
Given any quantum gate $A$ in (2.6.2) which operates on a single qubit, let us define for every $1 \leq i \leq n$ the
corresponding extension to an $n$-qubit quantum gate

\bigskip
(2.6.4) \quad $ A_i : {\bf C}^{2^n} ~\longrightarrow~ {\bf C}^{2^n} $

\medskip
which operates according to

\bigskip
(2.6.5) \quad $ \begin{array}{l} A_i ( |~ \psi_1 >, ~.~.~.~ , |~ \psi_n > ) ~=~ \\ \\
                                                         ~=~ ( |~ \psi_1 >, ~.~.~.~ , |~ \psi_{i - 1}, A |~ \psi_i >, |~ \psi_{i + 1}, ~.~.~.~ ,   |~ \psi_n > )
                        \end{array} $

\medskip
where $|~ \psi_1 >, ~.~.~.~ , |~ \psi_n > ~\in {\bf C}^2$. In other words, $A_i$ leaves all the qubits the same, except for
$|~ \psi_i >$, on which it operates according to the one qubit gate $A$. \\

Now, given $1 \leq i, j \leq n,~ i \neq j$, we extend the CNOT gate in (2.4.1), (2.4.2) to the following $n$-qubit gate

\bigskip
(2.6.6) \quad $ \mbox{CNOT}_{i, j} : {\bf C}^{2^n} ~\longrightarrow~ {\bf C}^{2^n} $

\medskip
which when applied to an arbitrary $n$-qubit  $( |~ \psi_1 >, ~.~.~.~ , |~ \psi_n > )$, leaves all the qubits the same, except
for $|~ \psi_i >$ and $|~ \psi_j  >$, upon which acts according to Fig 2.4.1. \\
It is easy to check that both $A_i$ and $\mbox{CNOT}_{i, j}$ defined above are unitary operators. \\

Then every $n$-qubit quantum gate $U$ in (2.6.3) can be written as the following decomposition

\bigskip
(2.6.7) \quad $ U ~=~ U_1 ~.~.~.~ U_m $

\medskip
for suitable $m \geq 1$ and with $U_1, ~.~.~.~ , U_m$ being either of the form (2.6.4) or (2.6.6). \\

In case we do {\it not} ask for equality, as in (2.6.7), and we are only looking for an {\it approximation} of $n$-qubit
quantum gates $U$ in (2.6.3), we have the following result which, on the other hand, is {\it stronger}, since it allows the
use of one {\it single} 2-qubit quantum gate. \\

Namely, given a 2-qubit quantum gate $B : {\bf C}^4 \longrightarrow {\bf C}^4$, we extend it to an $n$-qubit quantum gate

\bigskip
(2.6.8) \quad $ B_{i, j} : {\bf C}^{2^n} ~\longrightarrow~ {\bf C}^{2^n} $

\medskip
in a similar way as was done above for CNOT. \\

Then, there exist {\it universal} 2-qubit quantum gates $B$ such that for every $n$-qubit quantum gate $U$ and every
$\epsilon > 0$, one can find $1 \leq i_1, ~.~.~.~ , i_m, j_1, ~.~.~.~ , j_m \leq n$, with $i_1 \neq j_1, ~.~.~.~ , i_m \neq j_m$,
and with

\bigskip
(2.6.9) \quad $ | | ~ U ~-~ B_{i_1, j_1} ~.~.~.~ B_{i_m, j_m} ~| | ~\leq~ \epsilon $

\medskip
It is further known that a {\it generic} set of 2-qubit quantum gates $B : {\bf C}^4 \longrightarrow {\bf C}^4$ have the
above universal approximation property. In other words, this property is valid for an open and dense subset of such
quantum gates $B : {\bf C}^4 \longrightarrow {\bf C}^4$. However, when one is given a specific 2-qubit quantum gate, it
is not easy to check whether indeed it is universal in the above sense. \\

Let us conclude with a related result, and its proof, which can offer certain additional specifics, Deutsch [1]. Given

\bigskip
(2.6.10) \quad $ U : {\bf C}^D ~\longrightarrow~ {\bf C}^D $

\medskip
any {\it unitary} operator, where $D \geq 1$. Then there exists an orthonormal basis in ${\bf C}^D$ and unitary operators
$U_1, ~.~.~.~ , U_m : {\bf C}^D ~\longrightarrow~ {\bf C}^D$, with $m = 2 D^2 - D$, such that

\bigskip
(2.6.11) \quad $ U ~=~ U_1 ~.~.~.~ U_m $

\medskip
where each of the $U_i$ act on at most a two dimensional subspace ${\bf C}^D$ in the given basis. \\

Before presenting the proof of this property, let us note its consequence in the particular case of $n$-qubit quantum
gates $U$ in (2.6.3), when we have $D = 2^n$ and thus $m = 2^{2 n + 1} - 2^n$. Namely, every such quantum gate $U$
operating on $n$ qubits can be decomposed as in (2.6.11), where the $U_i$ are one qubit or two qubit quantum gates. \\

In order to show the general property (2.6.11), let $|~ \psi_1 >, ~.~.~.~ , |~ \psi_D > ~\in {\bf C}^D$ be the eigenvectors of the
unitary operator $U$, while $\lambda_1, ~.~.~.~ , \lambda_D \in {\bf C}$ denote the corresponding eigenvalues. \\
Given a certain fixed basis in ${\bf C}^D$, then $|~ \psi_1 >$ has in this basis the coordinates $( c_1, ~.~.~.~ , c_D )$. We
consider now the $D \times D$ block diagonal matrix

$$ A_{1, 2} ~=~ \left ( \begin{array}{l} \left ( \begin{array}{l} ~\bar{c}_1 / c_{1, 2} ~~~~~ ~\bar{c}_2 / c_{1, 2} \\ \\
                                                                                                      -c_2 / c_{1, 2} ~~~~~ c_1 / c_{1, 2}
                                                                    \end{array} \right )~~~
                                                            I_{3, ~.~.~.~ , D}
                                  \end{array} \right ) $$

\medskip
where $c_{1, 2} = ( | c_1 |^2 + | c_2 |^2 )^{1 / 2}$, while $I_{3, ~.~.~.~ , D}$ is the $( D - 2 ) \times ( D - 2 )$ identity
matrix. \\
Obviously $A_{1, 2}$ is unitary, and it operates on ${\cal C}^D$ only on the two dimensional subspace corresponding to
the first two coordinates in the given basis, and maps $|~ \psi_1 >$ into a vector with coordinates $( c_{1, 2}, 0, c_3,
~.~.~.~ , c_D )$. Applying further the similar matrices $A_{1, 3}, ~.~.~.~ , A_{1, D}$, one obtains a vector with coordinates
$( 1, 0, ~.~.~.~ , 0 )$. \\
Now we multiply the vector with coordinates $( 1, 0, ~.~.~.~ , 0 )$ with the eigenvalue $\lambda_1 = e^{i \theta_1}$, which
clearly is a unitary operator on ${\bf C}^D$ acting only on the one dimensional subspace corresponding to the first
coordinate in the given basis. \\
Further, in the order reverse to the one above, we apply the operators $A_{1, D}, ~.~.~.~ , A_{1, 2}$, and thus obtain
$\lambda_1~ |~ \psi_1 >$. \\
Obviously, we used for that purpose $2 D - 1$ unitary operators which acted on subspaces of dimension at most two. \\
Since the eigenvectors are orthogonal, the above procedure can be applied step by step to the other $D - 1$
eigenvectors, without disturbing the results obtained in previous steps. This then completes the proof of (2.6.11). \\

As mentioned, in quantum computers the decomposition (2.6.11) is of interest when $D = 2^n$ and thus $m = 2^{2 n + 1}
- 2^n$, where $n$ is the number of qubits on which the quantum gate $U$ in (2.6.3) operates. This however gives in
(2.6.11) a decomposition of $U$ which is {\it not} necessarily linear or even polynomial in the number $n$ of qubits on
which they operate. In this way, further improvements of such decompositions are useful. In this regard there are
known a number of results, and certain details can be found in Pittenger [pp. 24,25], Alber et.al. [pp. 98-109], as well as
the literature cited there. \\

It should be noted that the above results can have a {\it two fold} practical importance. Indeed, they allow the use of
very simple quantum gates when building up more complex quantum algorithms. Also, they may allow a convenient
architecture when building effective physically realized quantum computers. \\ \\

\chapter{Two Strange Phenomena}

We present next two novel and typical quantum computation phenomena. It is useful to encounter them early in the
study of quantum computation, since they can give an instructive insight into how much different quantum computers
are from the usual electronic digital ones. \\
The first we start with, called {\it no-cloning}, is an unexpected {\it limitation} in view of what we have been accustomed
to with usual electronic digital computers. The second one, called {\it teleportation}, is at least as surprising, however,
it can present great {\it advantages}. In this way, once again, we are in the situation described by "you win some, you
lose some ..." ... \\
Teleportation is also of interest since it makes essential use of quantum {\it entanglement} through double qubits in
${\bf C}^2 \otimes {\bf C}^2 \simeq {\bf C}^4$ which are called {\it EPR} or {\it  Bell pairs}. As far as no-cloning is
concerned, it proves to be an impossibility which results from very simple basic quantum principles. \\

{\bf 3.1 No-Cloning}

\bigskip
Scientists are on occasion giving names to new phenomena in ways which are not thoroughly enough considered, and
thus may lend themselves to misinterpretation. \\
One such case is with the term {\it no-cloning} used in quantum computation. \\
What is in fact going on here is that, quite surprisingly, quantum computers do {\it not} allow the copying of {\it arbitrary}
qubits. Thus a more proper term would be the somewhat longer one of {\it no arbitrary copying}. \\
Yet in spite of that, plenty of copying can be done by quantum computers, as will be seen in the sequel. \\

In order better to understand the issue, let us start by considering copying classical bits. For that purpose we can use
the classical version of the quantum CNOT gate in Fig. 2.4.1, operating this time on bits $a,~ b \in \{~ 0, 1 ~\}$, namely \\

\bigskip
\begin{math}
\setlength{\unitlength}{1cm}
\thicklines
\begin{picture}(15,3)
\put(2.8,2.8){$\mbox{a}$}
\put(3.5,2.9){\line(1,0){5}}
\put(9,2.8){$\mbox{a}$}
\put(2.8,0.1){$\mbox{b}$}
\put(3.5,0.2){\line(1,0){5}}
\put(9,0.1){$\mbox{a} ~\oplus~ \mbox{b}$}
\put(6,2.85){\circle*{0.5}}
\put(6,0.2){\line(0,1){2.6}}
\put(6,0.2){\circle{0.5}}
\put(5.1,-1){$\mbox{Fig. 3.1.1}$}
\end{picture}
\end{math} \\ \\

\medskip
Now, if we fix $b = 0$, then for an arbitrary input bit $a \in  \{~ 0, 1 ~\}$, we shall obtain as output two copies of $a$. \\

Strangely enough, a similar copying of arbitrary quantum bits cannot be performed by quantum systems, as was
discovered in 1982 by W K Wooters and W H Zurek, see Hirvensalo. \\
Of course, as seen in (2.1.2) - (2.1.4), each qubit contains a double infinity of classical information, much unlike the
situation with one single bit. In this way, the ability to copy arbitrary qubits is considerably more demanding than
copying arbitrary classical bits. \\

Let us now turn to this issue in some more detail. First we present a simple and somewhat intuitive argument. We
assume that we have a quantum system $S$ which allows one qubit at input and has one qubit at output. The output
facility we shall use as a "blank sheet" on which we want to copy an arbitrary input qubit $|~ \psi > ~\in {\bf C}^2$. We
can assume that the initial state of the "blank sheet" at the output is given by a fixed qubit $|~ \chi_0 > ~\in {\bf C}^2$.
Thus we start with the setup \\

\begin{math}
\setlength{\unitlength}{1cm}
\thicklines
\begin{picture}(15,3)
\put(2,1.4){$|~ \psi >$}
\put(3.5,1.5){\line(1,0){1.5}}
\put(5,0.5){\line(0,1){2}}
\put(5,0.5){\line(1,0){2}}
\put(5,2.5){\line(1,0){2}}
\put(7,0.5){\line(0,1){2}}
\put(7,1.5){\line(1,0){1.5}}
\put(9,1.4){$|~ \chi_0 >$}
\put(5.8,1.4){$S$}
\put(5.1,-0.5){$\mbox{Fig. 3.1.2}$}
\end{picture}
\end{math} \\ \\

\medskip
and would like to end up with the setup \\

\begin{math}
\setlength{\unitlength}{1cm}
\thicklines
\begin{picture}(15,3)
\put(2,1.4){$|~ \psi >$}
\put(3.5,1.5){\line(1,0){1.5}}
\put(5,0.5){\line(0,1){2}}
\put(5,0.5){\line(1,0){2}}
\put(5,2.5){\line(1,0){2}}
\put(7,0.5){\line(0,1){2}}
\put(7,1.5){\line(1,0){1.5}}
\put(9,1.4){$|~ \psi >$}
\put(5.8,1.4){$S$}
\put(5.1,-0.5){$\mbox{Fig. 3.1.3}$}
\end{picture}
\end{math} \\ \\

\medskip
However, as quantum processes evolve through unitary operators when not subjected to measurement, it means that
we are looking for such a unitary operator $U : {\bf C}^2 ~\otimes~ {\bf C}^2 ~\longrightarrow~ {\bf C}^2 ~\otimes~ {\bf C}^2$,
and one which would act according to

\bigskip
(3.1.1) \quad $ U (~ |~ \psi > ~\otimes~ |~ \chi_0 > ~) ~=~ |~ \psi > ~\otimes~ |~ \psi >,~~~~ |~ \psi > ~\in {\bf C}^2 $

\medskip
Before going further, let us immediately remark here that a unitary operator $U$, which therefore is linear, is not likely
to satisfy (3.1.1), in view of the fact that it is a {\it nonlinear}, in particular, quadratic relation in  $|~ \psi >
~\in {\bf C}^2$. \\

And now, let us return to a more precise argument. Since $|~ \psi > ~\in {\bf C}^2$ is assumed to be arbitrary in (3.1.1),
we can write that relation for any $|~ \psi_1 >,~|~ \psi_2 > ~\in {\bf C}^2$. Thus we obtain

\bigskip
(3.1.2) \quad $ \begin{array}{l} U (~ |~ \psi_1 > ~\otimes~ |~ \chi_0 > ~) ~=~ |~ \psi_1 > ~\otimes~ |~ \psi_1 > \\ \\
                                                  U (~ |~ \psi_2 > ~\otimes~ |~ \chi_0 > ~) ~=~ |~ \psi_2 > ~\otimes~ |~ \psi_2 >
                         \end{array} $

\medskip
Now if we take the inner product of these two relations and recall that $U$ was supposed to be unitary, we obtain

\bigskip
(3.1.3) \quad $ < \psi_1~ ~|~  \psi_2 > ~=~ (~ < \psi_1 ~|~  \psi_2 > ~)^2 $

\medskip
which implies that either $< \psi_1~ ~|~  \psi_2 > = 0$, or $< \psi_1~ ~|~  \psi_2 > = 1$. This means that the two arbitrary
quantum states $|~ \psi_1 >,~ |~ \psi_2 > ~\in {\bf C}^2$ are always either orthogonal, or identical from quantum point of
view, which is clearly absurd. \\

The general and rigorous argument is as follows. We consider a quantum system whose state space is ${\bf C}^n$, for
a certain integer $n \geq 1$. Further, we fix in this state space an arbitrary orthonormal basis $|~ \psi_1 >,~ .~.~.~ ,
|~ \psi_n > ~\in {\bf C}^n$. Finally, we assume that the state $|~ \psi_1 >~$ will function as the "blank sheet" on which we
want to copy arbitrary states $|~ \psi > ~\in {\bf C}^n$. \\
Then the desired copying machine of arbitrary states in ${\bf C}^n$ will be given by a unitary operator $U : {\bf C}^n
~\otimes~ {\bf C}^n ~\longrightarrow~ {\bf C}^n ~\otimes~ {\bf C}^n$, for which we have

\bigskip
(3.1.4) \quad $ U (~ |~ \psi > ~\otimes~ |~ \psi_1 > ~) ~=~ |~ \psi > ~\otimes~ |~ \psi >,~~~~ |~ \psi > ~\in {\bf C}^n $

\medskip
And now we can prove that for $n \geq 2$, there does {\it not} exist such a copying machine $U$. \\

Indeed, if we assume that $n \geq 2$, then we do have at least the two orthonormal states $|~ \psi_1 >,~ |~ \psi_2 > ~\in
{\bf C}^n$. Thus (3.1.4) gives

\bigskip
(3.1.5) \quad $ \begin{array}{l} U (~ |~ \psi_1 > ~\otimes~ |~ \psi_1 > ~) ~=~ |~ \psi_1 > ~\otimes~ |~ \psi_1 > \\ \\
                                                  U (~ |~ \psi_2 > ~\otimes~ |~ \psi_1 > ~) ~=~ |~ \psi_2 > ~\otimes~ |~ \psi_2 > \\ \\
                                                  U  (~ (~ |~ \psi_1 > ~+~ |~ \psi_2 > ~) ~\otimes~ |~ \psi_1 > ~) ~=~ \\
                                                        ~~~~~=~ (~ |~ \psi_1 > ~+~ |~ \psi_2 > ~) ~\otimes~ (~ |~ \psi_1 > ~+~ |~ \psi_2 > ~)
                         \end{array} $

\medskip
Now the last relation in (3.1.5) and the linearity of $U$ give together with the first two relations

\bigskip
(3.1.6) \quad $ \begin{array}{l} U  (~ (~ |~ \psi_1 > ~+~ |~ \psi_2 > ~) ~\otimes~ |~ \psi_1 > ~) ~=~ \\
                                                       ~~~~~ =~ U (~ |~ \psi_1 > ~\otimes~ |~ \psi_1 > ~) ~+~ U (~ |~ \psi_2 > ~\otimes~
                                                                                                                                                    |~ \psi_1 > ~) ~=~ \\
                                                       ~~~~~ =~ |~ \psi_1 > ~\otimes |~ \psi_1 > ~+~ |~ \psi_2 > ~\otimes |~ \psi_2 >
                         \end{array} $

\medskip
Thus (3.1.6) with the last relation in (3.1.5) imply

\bigskip
\hspace{2cm} $ (~ |~ \psi_1 > ~+~ |~ \psi_2 > ~) ~\otimes~ (~ |~ \psi_1 > ~+~ |~ \psi_2 > ~) ~=~ \\
                               ~~~~~~~~~~~~~~~~~~~~~~~ =~ |~ \psi_1 > ~\otimes |~ \psi_1 > ~+~ |~ \psi_2 > ~\otimes |~ \psi_2 > $

\medskip
or in other words

\bigskip
\hspace{2cm} $ |~ \psi_1 > ~\otimes~ |~ \psi_2 > ~+~ |~ \psi_2 > ~\otimes~ |~ \psi_1 > ~=~ 0 $

\medskip
which is obviously false. \\

Let us point out two facts with respect to the above no-cloning result. \\

First, in the more general second proof, we did not use the fact that $U$ is unitary, and only made use of its linearity,
when we obtained (3.1.6). In the first proof, on the other hand, the fact that $U$ is unitary was essential in order to
obtain (3.1.3). \\

Second, it is important to understand properly the meaning of the above limitation implied by no-cloning. Indeed, while
it clearly does not allow the copying of arbitrary qubits, it does nevertheless allow the copying of a {\it large range} of
qubits. \\
For instance, in terms of the second proof, let $I = \{~ 1, ~.~.~.~ , n ~\}$ be the set of indices of the respective orthonormal
basis

$$|~ \psi_1 >,~ .~.~.~ , |~ \psi_n > ~\in {\bf C}^n$$

\medskip
Further, let us consider the partially defined function

$$c : I \times I ~\longrightarrow~ I \times I$$

\medskip
given by $c~ ( i, 1 ) = ( i, i )$, with $1 \leq i \leq n$. Then clearly, $c$ is injective on the domain on which it is defined.
Therefore, $c$ can be extended to the whole of $I \times I$, so as still to remain injective, and in fact, become bijective.
And obviously, there are many such extensions when $n \geq 2$. \\
Now we can define a mapping $U$ by

$$U (~ |~ \psi_i > ~\otimes~ |~ \psi_j > ~) ~=~ |~ \psi_k > ~\otimes~ |~ \psi_l >$$

\medskip
where $1 \leq i,~ j \leq n$ and $c~ ( i, j ) = ( k, l )$. Since $c$ is bijective on $I \times I$, this mapping $U$ will be a
permutation of the respective basis in ${\bf C}^n ~\otimes~ {\bf C}^n$, therefore it extends in a unique manner to a linear
and unitary mapping

$$U : {\bf C}^n ~\otimes~ {\bf C}^n ~\longrightarrow~ {\bf C}^n ~\otimes~ {\bf C}^n$$

\medskip
And now it follows that

$$ U (~ |~ \psi_i > ~\otimes~ |~ \psi_1 > ~) ~=~ |~ \psi_i> ~\otimes~ |~ \psi_i >,~~~~~ 1 \leq i \leq n $$

\medskip
thus indeed $U$ is a copying machine with the "blank sheet" $|~ \psi_1 >$, and it can copy onto this "blank sheet" {\it
all} the qubits in the given orthonormal basis $|~ \psi_1 >,~ .~.~.~ , |~ \psi_n >~$ of ${\bf C}^n$. And in any such basis, with
the exception of the fixed "blank sheet" $|~ \psi_1 >$, all the other qubits $|~ \psi_2 >,~ .~.~.~ , |~ \psi_n >~$ are {\it
arbitrary}, within the constraint that together they have to form an orthonormal basis. \\ \\

{\bf 3.2 Teleportation}

\bigskip
The term {\it teleportation} used in the context of quantum computation is also somewhat misleading. Indeed, as we
shall see, there is no physical transportation of any kind taking place. What happens instead is that the specific
quantum state of a given input qubit $|~ \psi > ~=~ \alpha~ |~ 0 > ~+~ \beta~ |~ 1 > ~\in {\bf C}^2$ is reproduced identically
as an output. \\
This is however not copying either, since the input qubit will typically get destroyed in the process. More precisely, the
input qubit $|~ \psi >$ will be subjected to measurements which, in general, will therefore make it {\it collapse} into the
states $|~ 0 >~$ or $|~ 1 >$. In this way the nearest we may come to any sort of teleportation is that of the doubly infinite
classical {\it information content} in a quantum qubit, but in no way of any part of the effective quantum physical system
which may have supported that qubit at the input. \\

Quantum teleportation is not only a strange phenomenon, but it also has a variety of important applications in quantum
computing, and more generally, in the fast emerging theory of information processing through quantum systems. \\

One possible more convenient manner to present quantum teleportation is the familiar one which uses the personages
Alice and Bob who are supposed to be involved in this process. \\
The essential novel starting point in teleportation is that sometime in the past, Alice and Bob were together, generated
an entangled EPR pair, and then went apart, no matter how far, and for how long in time, from one another, each taking
with them one of the qubits from the entangled pair. \\

But let us first clarify the above by giving the following definition. An EPR pair is a double qubit in ${\bf C}^2 \otimes
{\bf C}^2 \simeq {\bf C}^4$ which has one of the following four forms

\bigskip
(3.2.1) \quad $ \begin{array}{l} |~ \omega_{~0 0} > ~=~ ( 1 / \sqrt 2 )~ (~ |~ 0 > ~\otimes~ |~ 0 > ~+~ |~ 1 > ~\otimes~
                                                                                                                                                                      |~ 1 > ~) \\ \\
                                                  |~ \omega_{~0 1} > ~=~ ( 1 / \sqrt 2 )~ (~ |~ 0 > ~\otimes~ |~ 1 > ~+~ |~ 1 > ~\otimes~
                                                                                                                                                                      |~ 0 > ~) \\ \\
                                                  |~ \omega_{~1 0} > ~=~ ( 1 / \sqrt 2 )~ (~ |~ 0 > ~\otimes~ |~ 0 > ~-~ |~ 1 > ~\otimes~
                                                                                                                                                                      |~ 1 > ~) \\ \\
                                                  |~ \omega_{~1 1} > ~=~ ( 1 / \sqrt 2 )~ (~ |~ 0 > ~\otimes~ |~ 1 > ~-~ |~ 1 > ~\otimes~
                                                                                                                                                                      |~ 0 > ~) \\ \\

                        \end{array} $

\medskip
Here the factor $1 / \sqrt 2$ is present in order to have the respective states $|~ \omega_{~i j} >~$ normalized in
${\bf C}^2 \otimes {\bf C}^2 \simeq {\bf C}^4$. In a simplified notation, which we shall use in the sequel, these four
quantum states will be written as

$$ \begin{array}{l} |~ \omega_{~0 0} > ~=~ ( 1 / \sqrt 2 )~ (~ |~ 0 0 > ~+~ |~ 1 1 > ~) \\ \\
                              |~ \omega_{~0 1} > ~=~ ( 1 / \sqrt 2 )~ (~ |~ 0 1 > ~+~ |~ 1 0 > ~) \\ \\
                              |~ \omega_{~1 0} > ~=~ ( 1 / \sqrt 2 )~ (~ |~ 0 0 > ~-~ |~ 1 1 > ~) \\ \\
                              |~ \omega_{~1 1} > ~=~ ( 1 / \sqrt 2 )~ (~ |~ 0 1 > ~-~ |~ 1 0 > ~)
     \end{array} $$

\medskip
There are three essential points to note with these EPR pairs. \\
First, they are entangled, since none of them is of the form $|~ \psi > ~\otimes~ |~ \chi >$, with $|~ \psi >,~ |~ \chi > ~\in {\bf
C}^2$. \\
Second, they belong to the composite quantum system ${\bf C}^2 \otimes {\bf C}^2$, thus Alice can take with her the
single qubit which belongs to the first factor ${\bf C}^2$ in this tensor product, while Bob can do the same with the
single qubit which belongs to the second factor. \\
Third, the entanglement means that, no matter how far, and for how long in time, Alice and Bob would go apart, there
will always be a certain nonclassical and typically quantum connection between their respective quantum qubits,
provided that, of course, none of the qubits is subjected to destruction. And according to Quantum Mechanics, this
quantum connection is not supposed to change or diminish with distance, or in time. \\

Needless to say, the above properties correspond not only to a thought experiment, but they can be effectively
implemented on suitable physically existent quantum systems. \\

For clarity, let us further note here that, in the case of the EPR pair $|~ \omega_{~0 0} >$, for instance, when Alice and
Bob take their respective single qubits from it, and then go apart, this does not at all refer to the terms $|~ 0 0 >$, or
$|~ 1 1 >$. Indeed, each of these two terms belongs to the composite quantum system ${\bf C}^2 \otimes {\bf C}^2$, and
thus to none of its separate two factors.Therefore they cannot be taken away either by Alice or by Bob. Needless to say,
the same goes for the other three EPR pairs as well. \\

The single qubits which Alice and Bob take with them respectively cannot be described in other way than it is already
done in the corresponding entangled states in (3.2.1), this being precisely one of the points about the typically quantum,
and nonclassical aspects of entanglement. \\

Of course, there are also more complicated, for instance, three or four term entangled quantum states in the composite
system ${\bf C}^2 \otimes {\bf C}^2$. Such examples are given, among many others, by the quantum states

$$ |~ 0 0 > ~+~ |~ 0 1 > ~+~ |~ 1 1 >,~~~ |~ 0 0 > ~+~ |~ 0 1 > ~-~ |~ 1 0 > ~+~ |~ 1 1 > $$

\medskip
However, the EPR pairs in (3.2.1), which have only two terms each, are some of the simplest possible entangled
quantum states, and as such, they can present certain advantages. \\
Needless to say, Alice and Bob could still take away their respective single qubits, regardless of the number of terms
in an entangled quantum state from the composite system ${\bf C}^2 \otimes {\bf C}^2$. \\

Let us now continue with the task Alice and Bob are facing when they are involved in quantum teleportation. Alice is
given a qubit $|~ \psi > ~=~ \alpha~ |~ 0 > ~+~ \beta~ |~ 1 > ~\in {\bf C}^2$. And she is not supposed to know it, since she is
not supposed to subject it to a measurement, which would typically risk to collapse it.. Yet by only using a classical
information channel with Bob, she has to let Bob obtain the full information about that qubit $|~ \psi >$. \\

At first sight, this seems to be an impossible task. Indeed, the qubit $|~ \psi >~$ contains a doubly infinite amount of
classical information, not to mention that Alice does not even have access to it. So that, even if Alice would fully know
the classical information contained in $|~ \psi >~$, she would not be in a position to convey it to Bob in finite time
through the classical information channel. \\

Fortunately, the task is nevertheless possible, due to the fact that Alice and Bob have kept intact their respective single
qubits from that entangled EPR pair which they had produced sometime in the past, when they were together. And the
task of the so called teleportation can be accomplished by the following device which is partly quantum and partly
classical

\begin{math}
\setlength{\unitlength}{1cm}
\thicklines
\begin{picture}(15,6)
\put(0,4.9){$|~ \psi >$}
\put(1.3,5){\line(1,0){1.5}}
\put(2.8,4.5){\line(0,1){1}}
\put(2.8,5.5){\line(1,0){1}}
\put(2.8,4.5){\line(1,0){1}}
\put(3.8,4.5){\line(0,1){1}}
\put(3.1,4.88){$\mbox{H}$}
\put(3.8,5){\line(1,0){1}}
\put(4.8,4.5){\line(0,1){1}}
\put(4.8,5.5){\line(1,0){1}}
\put(4.8,4.5){\line(1,0){1}}
\put(5.8,4.5){\line(0,1){1}}
\put(5.06,4.88){$\mbox{M}_1$}
\put(5.8,5.1){\line(1,0){3.6}}
\put(5.8,4.9){\line(1,0){3.4}}
\put(-0.5,1.9){$|~ \omega_{~0 0} >$}
\put(-0.2,2.9){$|~ \omega_A >$}
\put(1.3,3){\line(1,0){3.5}}
\put(4.8,2.5){\line(0,1){1}}
\put(4.8,3.5){\line(1,0){1}}
\put(4.8,2.5){\line(1,0){1}}
\put(5.8,2.5){\line(0,1){1}}
\put(5.06,2.88){$\mbox{M}_2$}
\put(5.8,3.1){\line(1,0){1.6}}
\put(5.8,2.9){\line(1,0){1.4}}
\put(2.05,5){\circle*{0.4}}
\put(2.05,3){\circle{0.4}}
\put(2.05,3){\line(0,1){2}}
\put(-0.2,0.9){$|~ \omega_B >$}
\put(1.3,1){\line(1,0){5.5}}
\put(6.8,0.5){\line(0,1){1}}
\put(6.8,1.5){\line(1,0){1}}
\put(6.8,0.5){\line(1,0){1}}
\put(7.8,0.5){\line(0,1){1}}
\put(7.4,3.1){\line(0,-1){1.6}}
\put(7.2,2.9){\line(0,-1){1.4}}
\put(7.1,0.88){$\mbox{X}_2$}
\put(7.8,1){\line(1,0){1}}
\put(8.8,0.5){\line(0,1){1}}
\put(8.8,1.5){\line(1,0){1}}
\put(8.8,0.5){\line(1,0){1}}
\put(9.8,0.5){\line(0,1){1}}
\put(9.4,5.1){\line(0,-1){3.6}}
\put(9.2,4.9){\line(0,-1){3.4}}
\put(9.1,0.88){$\mbox{Z}_1$}
\put(9.8,1){\line(1,0){1}}
\put(11.2,0.9){$|~ \psi >$}
\put(5.1,-0.5){$\mbox{Fig. 3.2.1}$}
\end{picture}
\end{math} \\ \\

\medskip
Here $|~ \psi >$ in the upper left corner is the input qubit at Alice which she wants to teleport to Bob, that is, to get to the
lower right output position. The other two inputs $|~ \omega_A >$ and $|~ \omega_B >$ are the entangled qubits in the
EPR pair $|~ \omega_{~0 0} >$, with the first of these qubits being at Alice, while the second one at Bob. \\
Further, $H$ is the Hadamard gate in (2.2.8), $X_2$ and $Z_1$ are certain adaptation to be specified of the Pauli gates
$X$ and $Z$, respectively, see (2.2.4), (2.2.6). The double lines are classical information channels, while $M_1,~ M_2$
are measuring devices specified later. \\

In order to follow the performance of the mixed quantum-classical device in Fig. 3.2.1, it is useful to break it up in four
successive input-output devices. The first of them is the following three qubit input, three qubit output quantum gate \\

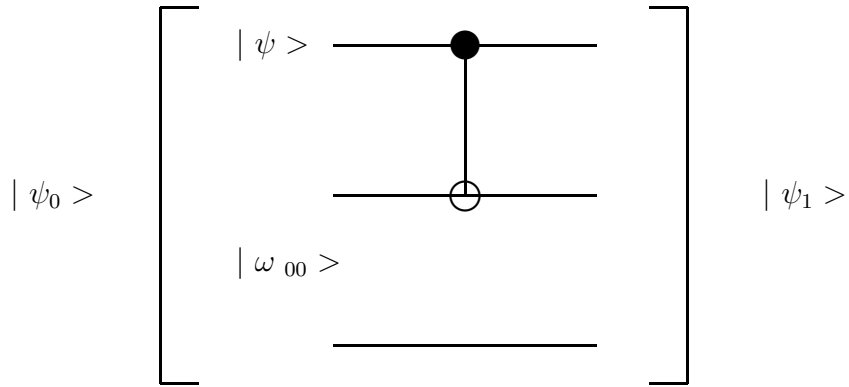
\begin{math}
\setlength{\unitlength}{1cm}
\thicklines
\begin{picture}(15,6)
\put(1,2.9){$|~ \psi_0 >$}
\put(3,5.5){\line(1,0){0.5}}
\put(3,5.5){\line(0,-1){5}}
\put(3,0.5){\line(1,0){0.5}}
\put(4,4.9){$|~ \psi >$}
\put(5.3,5){\line(1,0){3.5}}
\put(4,2){$|~\omega_{~0 0} >$}
\put(5.3,3){\line(1,0){3.5}}
\put(5.3,1){\line(1,0){3.5}}
\put(7.05,5){\circle*{0.4}}
\put(7.05,3){\circle{0.4}}
\put(7.05,3){\line(0,1){2}}
\put(10,5.5){\line(-1,0){0.5}}
\put(10,5.5){\line(0,-1){5}}
\put(10,0.5){\line(-1,0){0.5}}
\put(11,2.9){$|~ \psi_1 >$}
\end{picture}
\end{math} \\ \\

\medskip
in which we have the input

$$ |~ \psi_0 > ~=~ |~ \psi >~ |~ \omega_{~0 0} > ~=~ ~~~~~~~~~~~~~~~~~~~~~~~~~~~~~~~~~~~~~~~~~~~~~~~~~~~~~~~~~~~$$
$$ =~ ( 1 / \sqrt 2 )~(~ \alpha~ |~ 0 >~ (~ |~ 0 0 > ~+~ |~ 1 1 > ~) ~+~ \beta~ |~ 1 >~ (~ |~ 0 0 > ~+~ |~ 1 1 > ~) ~) $$

\medskip
In this three qubit input, the first two qubits, counted from the left, belong to Alice, while the first qubit counted from the
right belongs to Bob. In other words, Alice has the qubit $|~ \psi >$, as well as the left qubit from $|~ \omega_{~0 0} >$,
while Bob has the right qubit from $|~ \omega_{~0 0} >$. Let us now compute the three qubit output $|~ \psi_1 >$.
Clearly, Alice sends her two qubits through a CNOT gate, therefore

$$ |~ \psi_1 > ~=~ ~~~~~~~~~~~~~~~~~~~~~~~~~~~~~~~~~~~~~~~~~~~~~~~~~~~~~~~~~~~~~~~~~~~~~~~~~~~~~~~~~~~~$$
$$ =~ ( 1 / \sqrt 2 )~(~ \alpha~ |~ 0 >~ (~ |~ 0 0 > ~+~ |~ 1 1 > ~) ~+~ \beta~ |~ 1 >~ (~ |~ 1 0 > ~+~ |~ 0 1 > ~) ~) $$

\medskip
The second component of the device in Fig. 3.2.1 is again a three qubit input and three qubit output quantum gate,
namely

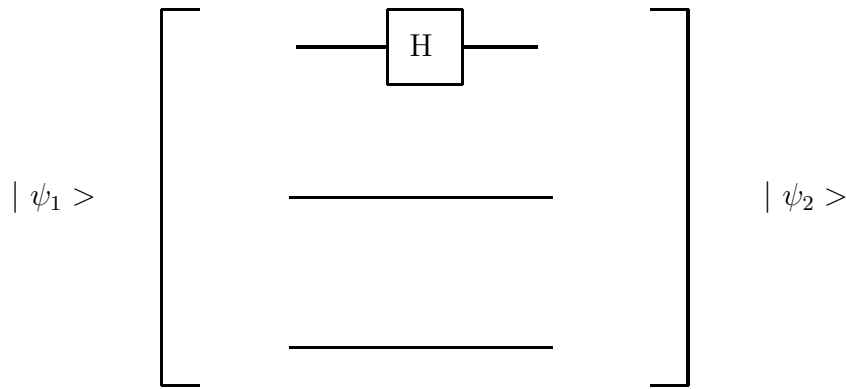
\begin{math}
\setlength{\unitlength}{1cm}
\thicklines
\begin{picture}(15,6)
\put(1,2.9){$|~ \psi_1 >$}
\put(3,5.5){\line(1,0){0.5}}
\put(3,5.5){\line(0,-1){5}}
\put(3,0.5){\line(1,0){0.5}}
\put(4.8,5){\line(1,0){1.2}}
\put(6,4.5){\line(0,1){1}}
\put(6,5.5){\line(1,0){1}}
\put(6,4.5){\line(1,0){1}}
\put(7,4.5){\line(0,1){1}}
\put(6.3,4.88){$\mbox{H}$}
\put(7,5){\line(1,0){1}}
\put(4.7,3){\line(1,0){3.5}}
\put(4.7,1){\line(1,0){3.5}}
\put(10,5.5){\line(-1,0){0.5}}
\put(10,5.5){\line(0,-1){5}}
\put(10,0.5){\line(-1,0){0.5}}
\put(11,2.9){$|~ \psi_2 >$}
\end{picture}
\end{math} \\ \\

\medskip
It follows that

$$ |~ \psi_2 > ~=~ ~~~~~~~~~~~~~~~~~~~~~~~~~~~~~~~~~~~~~~~~~~~~~~~~~~~~~~~~~~~~~~~~~~~~~~~~~~~~~~~$$
$$ =~ ( 1 / 2 )~ (~ \alpha~ (~ |~ 0 > ~+~ |~ 1 > ~)~(~ |~ 0 0 > ~+~ |~ 1 1 > ~) ~+~ ~~~~~~~~~~~~~~~~~~~~~~~~~~~$$
$$ +~ \beta~ (~ |~ 0 > ~-~ |~ 1 > ~)~(~ |~ 1 0 > ~+~ |~ 0 1 > ~) ~) $$

\medskip
By using the associativity of the tensor product, we further obtain

$$ |~ \psi_2 > ~=~ ~~~~~~~~~~~~~~~~~~~~~~~~~~~~~~~~~~~~~~~~~~~~~~~~~~~~~~~~~~~~~~~~~~~~~~~~~~~~~~~$$
$$ =~ ( 1 / 2 )~ (~ |~ 0 0 >~(~ \alpha~ |~ 0 > ~+~ \beta~ |~ 1 > ~) ~+~ |~ 0 1 >~(~ \alpha~ |~ 1 > ~+~ \beta~ |~ 0 > ~) ~+~ $$
$$ +~ ( 1 / 2 )~ (~ |~ 1 0 >~(~ \alpha~ |~ 0 > ~-~ \beta~ |~ 1 > ~) ~+~ |~ 1 1 >~(~ \alpha~ |~ 1 > ~-~ \beta~ |~ 0 > ~) ~+~ $$

\medskip
The expression in the right hand side is quite useful. Its first term

$$ |~ 0 0 >~(~ \alpha~ |~ 0 > ~+~ \beta~ |~ 1 > ~) $$

\medskip
has the two qubits of Alice in the state $|~ 0 0 >$ and the single cubit of Bob in the state $\alpha~ |~ 0 > ~+~ \beta~ |~ 1 >~$
which is in fact $|~ \psi >$. Therefore, if Alice performs a measurement on her two qubits at obtains $|~ 0 0 >$, then Bob
will have obtained the desired $|~ \psi >$. Proceeding in a similar fashion, we obtain the following table in which the left
column lists the four possible double bits of classical information which Alice can obtain by measuring her two qubits,
while the right column contains the corresponding states of the single qubit which Bob will obtain following the
measurement

\bigskip
(3.2.2) \quad $ \begin{array}{l} |~ 0 0 > ~~~\longrightarrow~~~ \alpha~ |~ 0 > ~+~ \beta~ |~ 1 > \\
                                                 |~ 0 1 > ~~~\longrightarrow~~~ \alpha~ |~ 1 > ~+~ \beta~ |~ 0 > \\
                                                 |~ 1 0 > ~~~\longrightarrow~~~ \alpha~ |~ 0 > ~-~ \beta~ |~ 1 > \\
                                                   |~ 1 1 > ~~~\longrightarrow~~~ \alpha~ |~ 1 > ~-~ \beta~ |~ 0 >
                        \end{array} $

\medskip
This leads to the third component of the device in Fig. 3.2.1 which this time is a mixed classical-quantum device with
three qubits as input, while its output are two classical bits and one qubit  \\

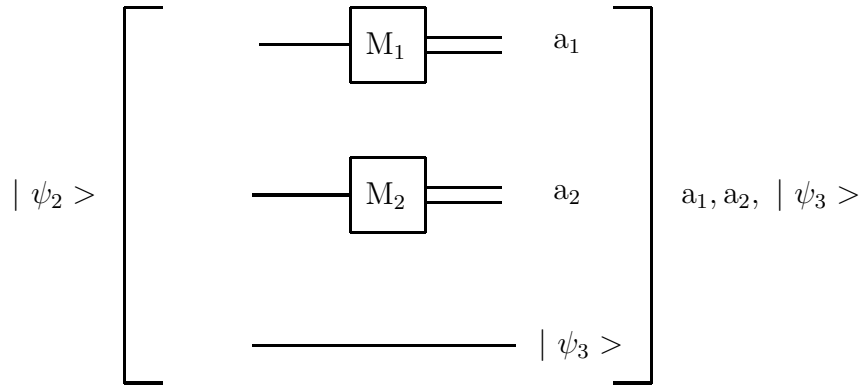
\begin{math}
\setlength{\unitlength}{1cm}
\thicklines
\begin{picture}(15,6)
\put(1.5,2.9){$|~ \psi_2 >$}
\put(3,5.5){\line(1,0){0.5}}
\put(3,5.5){\line(0,-1){5}}
\put(3,0.5){\line(1,0){0.5}}
\put(4.8,5){\line(1,0){1.2}}
\put(6,4.5){\line(0,1){1}}
\put(6,5.5){\line(1,0){1}}
\put(6,4.5){\line(1,0){1}}
\put(7,4.5){\line(0,1){1}}
\put(6.2,4.88){$\mbox{M}_1$}
\put(7,5.1){\line(1,0){1}}
\put(7,4.9){\line(1,0){1}}
\put(8.7,4.95){$\mbox{a}_1$}
\put(4.7,3){\line(1,0){1.3}}
\put(7,3.1){\line(1,0){1}}
\put(7,2.9){\line(1,0){1}}
\put(8.7,2.95){$\mbox{a}_2$}
\put(4.7,1){\line(1,0){3.5}}
\put(8.5,0.9){$|~ \psi_3 >$}
\put(6,2.5){\line(0,1){1}}
\put(6,3.5){\line(1,0){1}}
\put(6,2.5){\line(1,0){1}}
\put(7,2.5){\line(0,1){1}}
\put(6.2,2.88){$\mbox{M}_2$}
\put(10,5.5){\line(-1,0){0.5}}
\put(10,5.5){\line(0,-1){5}}
\put(10,0.5){\line(-1,0){0.5}}
\put(10.4,2.9){$\mbox{a}_1, \mbox{a}_2,~|~ \psi_3 >$}
\end{picture}
\end{math}

\medskip
Now the measurements $M_1$ and $M_2$ made by Alice will give her the bits $a_1$ and $a_2$, respectively. This is
precisely the classical information which she has to communicate to Bob. \\
And then based on table (3.2.2), Bob is at last in the position to receive the original qubit $|~ \psi > ~=~ \alpha~ |~ 0 >
~+~ \beta~ |~ 1 >$. For that purpose he can use the following mixed classical-quantum device with input two bits and a
qubit, and output one qubit, a device which is the fourth component of the device in Fig. 3.2.1 \\

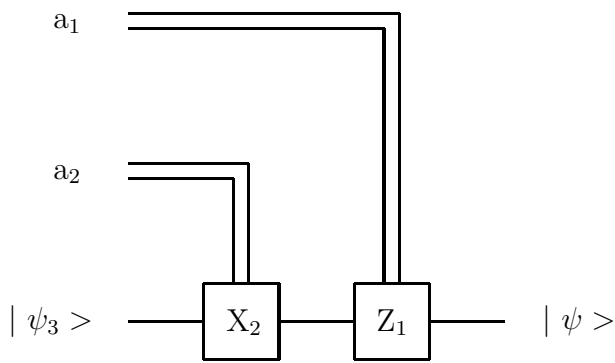
\begin{math}
\setlength{\unitlength}{1cm}
\thicklines
\begin{picture}(15,6)
\put(2,4.9){$\mbox{a}_1$}
\put(3,5.1){\line(1,0){3.6}}
\put(3,4.9){\line(1,0){3.4}}
\put(2,2.9){$\mbox{a}_2$}
\put(3,3.1){\line(1,0){1.6}}
\put(3,2.9){\line(1,0){1.4}}
\put(1.4,0.9){$|~ \psi_3 >$}
\put(3,1){\line(1,0){1}}
\put(4,0.5){\line(0,1){1}}
\put(4,1.5){\line(1,0){1}}
\put(4,0.5){\line(1,0){1}}
\put(5,0.5){\line(0,1){1}}
\put(4.6,3.1){\line(0,-1){1.6}}
\put(4.4,2.9){\line(0,-1){1.4}}
\put(4.3,0.88){$\mbox{X}_2$}
\put(5,1){\line(1,0){1}}
\put(6,0.5){\line(0,1){1}}
\put(6,1.5){\line(1,0){1}}
\put(6,0.5){\line(1,0){1}}
\put(7,0.5){\line(0,1){1}}
\put(6.6,5.1){\line(0,-1){3.6}}
\put(6.4,4.9){\line(0,-1){3.4}}
\put(6.3,0.88){$\mbox{Z}_1$}
\put(7,1){\line(1,0){1}}
\put(8.5,0.9){$|~ \psi >$}
\end{picture}
\end{math} \\ \\

\medskip
Here what happens is as follows. If $ a_1~ a_2 = 0 ~0$ then Bob already has $|~ \psi >$ as the output. If $a_1~ a_2 =
1~0$ then the gate $Z$ has  to be activated in order to obtain the same output. Further, in case $a_1~a_2 = 0~1$ then
the gate $X$ should be activated for obtaining again the desired output. Finally, when $a_1~a_2 = 1~1$, then both gates
$X$ and $Z$ have to be activated in this order, so that the output will be $|~ \psi >$. \\

Having analyzed in some detail the device in Fig. 3.2.1 used for quantum teleportation, one more observation can be
useful. Namely, the left-to-right direction, according to which by convention the flow of information is supposed to
happen in such diagrams, need not at the same time represent as well the effective spatial disposition of inputs,
outputs or other entities related to the respective process. This can be seen quite clearly even in the case of the
diagram in Fig. 3.2.1. Namely, the respective spatial disposition has, of course, part of this diagram located at Alice,
while the other part may be ways far away, at Bob. And this spatial separation is indicated by the following starred line
which divides this diagram in two, with the upper part being at Alice, and the lower part at Bob

\begin{math}
\setlength{\unitlength}{1cm}
\thicklines
\begin{picture}(15,6)
\put(0,4.9){$|~ \psi >$}
\put(1.3,5){\line(1,0){1.5}}
\put(2.8,4.5){\line(0,1){1}}
\put(2.8,5.5){\line(1,0){1}}
\put(2.8,4.5){\line(1,0){1}}
\put(3.8,4.5){\line(0,1){1}}
\put(3.1,4.88){$\mbox{H}$}
\put(3.8,5){\line(1,0){1}}
\put(4.8,4.5){\line(0,1){1}}
\put(4.8,5.5){\line(1,0){1}}
\put(4.8,4.5){\line(1,0){1}}
\put(5.8,4.5){\line(0,1){1}}
\put(5.06,4.88){$\mbox{M}_1$}
\put(5.8,5.1){\line(1,0){3.6}}
\put(5.8,4.9){\line(1,0){3.4}}
\put(-0.5,1.9){$|~\omega_{~0 0} >$}
\put(-0.2,2.9){$|~ \omega_A >$}
\put(1.2,1.9){$********************************$}
\put(1.3,3){\line(1,0){3.5}}
\put(4.8,2.5){\line(0,1){1}}
\put(4.8,3.5){\line(1,0){1}}
\put(4.8,2.5){\line(1,0){1}}
\put(5.8,2.5){\line(0,1){1}}
\put(5.06,2.88){$\mbox{M}_2$}
\put(5.8,3.1){\line(1,0){1.6}}
\put(5.8,2.9){\line(1,0){1.4}}
\put(2.05,5){\circle*{0.4}}
\put(2.05,3){\circle{0.4}}
\put(2.05,3){\line(0,1){2}}
\put(-0.2,0.9){$|~ \omega_B >$}
\put(1.3,1){\line(1,0){5.5}}
\put(6.8,0.5){\line(0,1){1}}
\put(6.8,1.5){\line(1,0){1}}
\put(6.8,0.5){\line(1,0){1}}
\put(7.8,0.5){\line(0,1){1}}
\put(7.4,3.1){\line(0,-1){1.6}}
\put(7.2,2.9){\line(0,-1){1.4}}
\put(7.1,0.88){$\mbox{X}_2$}
\put(7.8,1){\line(1,0){1}}
\put(8.8,0.5){\line(0,1){1}}
\put(8.8,1.5){\line(1,0){1}}
\put(8.8,0.5){\line(1,0){1}}
\put(9.8,0.5){\line(0,1){1}}
\put(9.4,5.1){\line(0,-1){3.6}}
\put(9.2,4.9){\line(0,-1){3.4}}
\put(9.1,0.88){$\mbox{Z}_1$}
\put(9.8,1){\line(1,0){1}}
\put(11.2,0.9){$|~ \psi >$}
\put(5.1,-0.5){$\mbox{Fig. 3.2.2}$}
\end{picture}
\end{math} \\ \\

\medskip
Clearly, Alice can only input the qubit $|~ \psi >~$ which is in the upper left corner, as well as her qubit $|~ \omega_A >~$
from the entangled EPR pair $|~ \omega_{~0 0} >$. On the other hand, Bob can only input his qubit $|~ \omega_B >~$
from the same pair. And although in the diagram the inputs are all on the left, it is nevertheless obvious that they are
far from being at the same place, at least not in the case of Alice and Bob in the above situation. \\

Finally, we should note that during teleportation as performed above, both the original qubit $|~ \psi >~$ and the
entangled EPR pair $|~ \omega_{~0 0} >~$ will in general become destroyed. Indeed, as mentioned, the original qubit
$|~ \psi >~$ is subjected to measurement, and this happens when it goes from the stage $|~ \psi_2 >~$ to the stage
$|~ \psi_3 >$, thus it suffers a {\it collapse}. The same happens with the qubit $|~ \omega_A >~$ of Alice, which is her
part of the EPR pair. \\

In this way, teleportation has a price, and a nontrivial quantum one at that : \\

One qubit teleported costs in general one entangled EPR pair ! \\ \\

\chapter{Bell's Inequalities}

We have seen some of the importance of the typically quantum phenomenon of {\it entanglement} when we used
entangled EPR pairs in quantum teleportation. This issue of entanglement has been, and still is of special focus in
Quantum Mechanics, not least due to its intimate connection to such fundamental disputes as {\it locality} versus {\it
nonlocality}. And as mentioned, the related literature is indeed vast. \\

Nearly three decades after the EPR paper had appeared in 1935, John Bell published in 1964 what amounted to a
surprising {\it conflict} between predictions of a classical world view based on the principle of locality, and on the other
hand, of Quantum Mechanics. The classical world view based on locality led J Bell to certain inequalities which,
however, proved to be {\it contradicted} by Quantum Mechanics, namely, by certain properties of suitably chosen
entangled EPR pairs. \\
And this contradiction could be observed in effective quantum mechanical experiments, such as conducted for instance
in 1982 by A Aspect et.al., see Maudlin. \\

Here it should be mentioned again that, as often, the related terminology which entered the common use tends to
misplace the focus. Indeed, the main point in J Bell's contribution is not about inequalities, but about the fact that they
lead to the mentioned contradiction. Furthermore, there are by now a number of other similar arguments which all lead
to such contradictions with Quantum Mechanics. \\

Needless to say, it is well known that ever since its very inception in the 1920s, Quantum Mechanics has been
witnessing an ongoing foundational controversy related to its interpretation, some of the earlier major stages of this
controversy being those between N Bohr and A Einstein. However, as not seldom in such human situations, a certain
saturation, stationarity and loss of interest may set in after some longer period of time has failed to clarify enough the
issues involved. \\

The surprising result of J Bell happened to appear after most of the founding fathers of Quantum Mechanics had left the
scene, and proved to inaugurate a fresh line of controversies, see Bell, Cushing \& McMullin, Maudlin. \\

Here, an attempt is presented to recall in short the essential aspects of J Bell's result. Clearly, at least to the extent
that this result is essentially connected to the typically quantum phenomenon of entanglement, it may be expected to
be relevant for a better understanding, and thus further development of quantum computation. \\

Also, a relatively less well know aspect of Bell type inequalities is presented here, namely that, these inequalities are
among a larger class of purely probabilistic inequalities, a class whose study was started by George Boole, with the
first results published in his book The Laws of Thought, back in 1854. This purely mathematical study was later further
extended in the work of a number of mathematicians and probabilists, see details Pitowsky, for instance. \\
Needless to say, this fact does in no way detract from the importance and merit of J Bell's result. Indeed, unlike J Bell,
it is obvious that G Boole and his mentioned followers, including those in more recent times, did not consider the
quantum mechanical implications of such inequalities. In this way, the importance and merit of J Bell's result is to
single out for the first time certain rather simple inequalities which are supposed to be universally valid, provided that
a classical setup and locality are assumed, and then show that the respective inequalities do to a quite significant
extent conflict with Quantum Mechanics, involving in this process such important issues as entanglement and locality
versus nonlocality. \\

There is a special interest in pointing out the fact that the Bell type inequalities can be established by a purely
mathematical argument, as was done, for instance, by the followers of G Boole. Indeed, both in the work of J Bell, as
well as in the subsequent one of many of the physicists who dealt with this issue, the true nature of such inequalities
is often quite obscured by a complicated mix of physical and mathematical argument. Such an approach, however, is
unnecessary, and can of course create confusions about the genuine meaning, scope and implications of J Bell's
result. \\

The fact however is that regardless of the considerable generality of the framework underlying such inequalities, and
thus of the corresponding minimal conditions required on locality, one can nevertheless obtain the respective
inequalities through purely mathematical argument, and {\it without} any physical considerations involved, yet they
turn out even to be testable {\it empirically}. And in a surprising manner, they fail tests which are of a quantum
mechanical nature. And this failure is both on theoretical and empirical level. In other words, the Bell inequalities
contradict theoretical consequences of Quantum Mechanics, and on top of that, they are also proven wrong in quantum
mechanical experiments such as those conducted by Aspect et.al. \\

The impact of Bell's inequalities is only increased by the fact that they require such {\it minimal} conditions, yet they
deliver a clear cut and unavoidable {\it conflict} with Quantum Mechanics. \\

Let us also note the following. J Bell, when obtained his inequalities, he was concerned with the issue of the possibility, or
otherwise, of the so called {\it deterministic, hidden variable} theories for Quantum Mechanics. This issue arose from
the basic controversy in the interpretation of Quantum Mechanics, and aimed to eliminate the probabilistic aspects
involved typically in the outcome of measurements. One way in this regard was to consider Quantum Mechanics {\it
incomplete}, and then add to it the so called hidden variables, thus making the theory deterministic by being able a
priori to specify precisely the measurement results. \\

By the way, the very title of the EPR paper was raising the question whether Quantum Mechanics was indeed complete,
and suggested the experiment with entangled quantum states in order to justify that questioning. \\

Regarding the term hidden variables, once again we are faced with a less than proper terminology. Indeed, as it is
clear from the context in which this term has always been used, one is rather talking about {\it missing variables}, or
perhaps variables which have been missed, overlooked or disregarded, when the theory of Quantum Mechanics
was set up. Details in this regard can be found in Holland, where an account of the de Broglie-Bohm causal approach
to Quantum Mechanics is presented. \\

In view of this historical background, the effect of Bell's inequalities is often {\it wrongly} interpreted as proving that a
deterministic hidden variable theory which is subjected to the principle of locality is not possible. \\

However, it is important to note that such a view of Bell's inequalities is {\it not} correct. Indeed, by giving up
determinism, or the hidden variables, one still remains with Bell's inequalities, since these inequalities {\it only}
assume a classical framework in which the locality principle holds. \\ \\

{\bf 4.1 Boole Type Inequalities}

\bigskip
In his mentioned book G Boole was concerned among others with conditions on all possible experience or
experimentation, this being the factual background to logic and the laws of thought. Needless to say, G Boole assumed
automatically a classical and non-quantum context which was further subjected to the principle of locality. \\

Here we shall limit ourselves to a short presentation of some of the relevant aspects. Let therefore $A_1, ~.~.~.~ , A_n$
be arbitrary $n \geq 2$ events, and for $1 \leq i_1 < i_2 < ~.~.~.~ < i_k \leq n$, let $p_{i_1, i_2,  ~.~.~.~ , i_k}$ be the
probability of the simultaneous event $A_{i_1} \bigcap A_{i_2} \bigcap ~.~.~.~ \bigcap A_{i_k}$. \\

One of the questions G Boole asked was as follows. Suppose that the only information we have are
the probabilities
$p_1,~ p_2, ~.~.~.~ , p_n$ of the respective individual events $A_1, ~.~.~.~ , A_n$. What are under these
conditions on information the best possible
estimates for the probabilities of $A_1 \bigcup A_2 \bigcup ~.~.~.~ \bigcup A_n$ and $A_1
\bigcap A_2 \bigcap ~.~.~.~
\bigcap A_n$ ? \\

G Boole gave the following answers which indeed are correct

\bigskip
(4.1.1) \quad $ \begin{array}{l} \mbox{max}~ \{~ p_1,~ p_2, ~.~.~.~ , p_n ~\} ~\leq~ P (  A_1 \bigcup A_2 \bigcup ~.~.~.~
                                                    \bigcup A_n ) ~\leq~ \\ \\
                                                        ~~~~~~~~~~~~~~~~~~~~~~~~~~~~~~~~~~~~~~~~~~\mbox{min}~ \{~ 1,~ p_1 + p_2 + ~.~.~.~  + p_n ~\}
                        \end{array} $

\medskip
(4.1.2) \quad $ \begin{array}{l} \mbox{max}~ \{~ 0,~ p_1 + p_2 + ~.~.~.~ + p_n -n + 1 ~\} ~\leq~ \\ \\
                                                  ~~~~~~~~~\leq~ P (  A_1 \bigcap A_2 \bigcap  ~.~.~.~ \bigcap A_n ) ~\leq~
                                                                                            ~\mbox{min}~ \{~  p_1 ,~ p_2, ~.~.~.~  , p_n ~\}
                         \end{array} $

\bigskip
And these are the best possible inequalities in general, since for suitable particular cases equality can hold in each of
the four places. \\

A rather general related result is the so called inclusion-exclusion principle of Henri Poincar\'{e}

\bigskip
(4.1.3) \quad $ \begin{array}{l} P (  A_1 \bigcup A_2 \bigcup ~.~.~.~  \bigcup A_n ) ~=~ \Sigma_{1 \leq i \leq n}~ p_i ~-~
                                                  \Sigma_{1 \leq i < j \leq n}~ p_{i j} ~+~ \\ \\
                                      ~~~~~~~~~~~~~~~+~ \Sigma_{1 \leq i < j < k \leq n}~ p_{i j k} ~+~ ~.~.~.~  +~ ( - 1 )^{n + 1}~ p_{1 2 ~.~.~. n}

                        \end{array} $

\medskip
This however requires the knowledge of the probabilities of all the simultaneous events $A_{i_1} \bigcap A_{i_2}
\bigcap  ~.~.~.~ \bigcap A_{i_k}$, with $1 \leq i_1 < i_2 < ~.~.~.~ < i_k \leq n$. \\

A question with less demanding data, yet with more of them than required in (4.1.1) and (4.1.2), is the following.
Suppose we know the probabilities $p_i$ of the events $A_i$, with $1 \leq i \leq n$, as well as the probabilities $p_{i j}$
of the simultaneous events $A_i \bigcap A_j$, with $1 \leq i < j \leq n$. \\
What is then the best possible estimate for the probability of \\ $A_1 \bigcup A_2 \bigcup ~.~.~.~  \bigcup A_n$ ? \\

Unfortunately, this question is computationally intractable, Pitowsky. However, C E Bonferroni gave some answers in
1936, one of which is that

\bigskip
(4.1.4) \quad $ \Sigma_{1 \leq i \leq n}~ p_i ~-~ \Sigma_{1 \leq i < j \leq n}~ p_{i j} ~\leq~ P (  A_1 \bigcup A_2 \bigcup ~.~.~.~
                                                    \bigcup A_n ) $

\medskip
and here it is interesting to note that (4.1.4) generates easily $2^n - 1$ other independent inequalities by the following
procedure. We take any $1 \leq i_1 < i_2 < ~.~.~.~ < i_k \leq n$, and replace in (4.1.4) the events $A_{i_l}$, for $1 \leq l
\leq k$, with their complementaries. \\

Now the important fact to note is that Bell's inequalities result from (4.1.4) in this way, in the case of $n = 3$. \\

It follows therefore that Bell's inequalities are of a purely mathematical nature, and as such, only depend on classical
probability theory. \\

By the way, Boole's inequalities and its further developments have been presented in well known monographs of
mathematics and probability theory, some of them as recently as in 1970, and related research has continued in
mathematics and in probability theory till the present day, Pitowsky. As so often however, due to extreme
specialization and the corresponding narrowing of interest, such results seem not to be familiar among quantum
physicists. In this regard it may be worth mentioning that Pitowski himself is a philosopher of science. \\ \\

{\bf 4.2 The Bell Effect}

\bigskip
There are by now known a variety of ways which describe the phenomenon brought to light for the first time by Bell's
inequalities. In order to avoid complicating the issues involved, we shall present here one of the most simple such
ways, Maudlin. \\
This phenomenon, which one can call the {\it Bell effect} is a {\it contradiction} resulting between Quantum Mechanics,
and on the other hand, what can be done in a classical setup which satisfies the principle of locality. The Bell
inequalities are only one of the ways, and historically the first, which led to such a contradiction. They will be
presented in section 4.3. What is given here is a simple and direct argument leading to the mentioned kind of
contradiction. \\

Certain {\it entangled} quantum particles can exhibit the following behaviour. After they become spatially separated,
they each can be subjected to three different experiments, say, A, B and C, and each of them can produce one and only
one of two results, which for convenience we shall denote by R and S, respectively. \\
What is so uniquely specific to these entangled quantum particles is the behaviour described in the next three
conditions which such particles do satisfy. \\

{\it Condition 1}. When both particles are subjected to the same experiment, they give the same result. \\

{\it Condition 2}. When one of the particles is subjected to A and the other to B, or one is subjected to B and the other to
C, they will in a large number of experiments give the same result with a frequency of 3/4. \\

{\it Condition 3}. When one of the particles is subjected to A and the other to C, then in a large number of experiments
they will give the same result with a frequency 1/4. \\

Now, the surprising fact is that {\it no} experiment in a classical setup in which the principle of locality holds can come
anywhere near to such a behaviour. \\
And strangely enough, that includes as well the case when two conscious participants, and not merely two physical
entities would be involved. In such a case, when conscious participant are present, we shall see the experiments A, B
and C as questions put to the two participants, while the results R and C will be seen as their respective answers. \\

Such are indeed the wonders of {\it entanglement} and of certain EPR pairs that some of their performances, like for
instance those which satisfy conditions 1, 2 and 3 above, {\it cannot} be reproduced in a classical context which obeys
the locality principle, even if attempted by two conscious participants. \\

Indeed, a simple analysis shows that the best two such participants can do is to decide to give the same answers,
when asked the same questions. This means that any possible {\it strategy} of the two participants has to be {\it joint}
or identical, and as such, it is given by a function $f : \{~ A, B, C ~\} \longmapsto \{~ R, S ~\}$. \\
Clearly, there are 8 such joint strategies, namely

\bigskip
(4.2.1) \quad $ \begin{array}{l} ~~~~~~~~~ A ~~~~~~ B ~~~~~~ C \\
                                                  ---------- \\ \\
                                                  1 ~~~~~~~ R ~~~~~~ R ~~~~~~ R \\
                                                  2 ~~~~~~~ R ~~~~~~ R ~~~~~~ S \\
                                                  3 ~~~~~~~ R ~~~~~~ S ~~~~~~ R \\
                                                  4 ~~~~~~~ R ~~~~~~ S ~~~~~~ S \\
                                                  5 ~~~~~~~ S ~~~~~~ R ~~~~~~ R \\
                                                  6 ~~~~~~~ S ~~~~~~ R ~~~~~~ S \\
                                                  7 ~~~~~~~ S ~~~~~~ S ~~~~~~ R \\
                                                  8 ~~~~~~~ S ~~~~~~ S ~~~~~~ S
                        \end{array} $ \\

\medskip
Now it is obvious that by choosing only these 8 join strategies, condition 1 above will be satisfied. \\
From the point of view of satisfying conditions 2 and 3 above, the strategy pairs ( 1, 8 ), ( 4, 5 ), ( 3, 6 ) and ( 2, 7 )
are equivalent. Therefore, we only remain with four distinct strategies to consider, namely

\bigskip
(4.2.2) \quad $ \begin{array}{l} ~~~~~~~~~ A ~~~~~~ B ~~~~~~ C \\
                                                  ---------- \\ \\
                                                  1 ~~~~~~~ R ~~~~~~ R ~~~~~~ R \\
                                                  2 ~~~~~~~ R ~~~~~~ R ~~~~~~ S \\
                                                  3 ~~~~~~~ R ~~~~~~ S ~~~~~~ R \\
                                                  4 ~~~~~~~ R ~~~~~~ S ~~~~~~ S \\
                         \end{array} $ \\

\medskip
At this point the two participants can further improve on their attempt to satisfy conditions 2 and 3 above by
randomizing their joint strategies. For that purpose, they can choose four real numbers $\alpha, \beta, \gamma, \delta
\in {\bf R}$, such that

\bigskip
(4.2.3) \quad $ \begin{array}{l} \alpha,~ \beta,~ \gamma,~ \delta ~\geq~ 0 \\ \\
                                                  \alpha ~+~\beta ~+~ \gamma ~+~ \delta ~=~ 1
                        \end{array} $

\medskip
and use their joint strategies 1, 2, 3 and 4 with the respective frequencies $\alpha, \beta, \gamma, \delta$. A simple
computation will show that conditions 2 and 3 above will further impose on $\alpha, \beta, \gamma, \delta$ the relations

\bigskip
(4.2.4) \quad $ \begin{array}{l} \gamma ~+~ \delta ~=~ 1/4 \\ \\
                                                  \beta ~+~ \gamma ~=~ 1/4 \\ \\
                                                  \beta ~+~ \delta ~=~ 3/4
                        \end{array} $

\medskip
However, (4.2.3) and (4.2.4) yield

\bigskip
(4.2.5) \quad $ \gamma ~=~ - 1/8 $

\medskip
thus a {\it contradiction}. \\

Here it is important to note that the {\it locality} principle was assumed in (4.2.1) - (4.2.5). In other words, each of the two
participants could be asked questions, without the question asked from one of them having any effect on the answer of
the other. Indeed, the two participants could be asked different questions, and each of them would only reply according
to the question asked, and according to their joint strategy, which they happened to use at the moment. \\

The fact that the setup in (4.2.1) - (4.2.5) is {\it classical}, that is, it is not specifically quantum mechanical, is
obvious. \\ \\

{\bf 4.3 Bell's Inequalities}

\bigskip
For convenience we shall consider two entangled quantum particles which are in a situation even simpler than in
section 4.2, Cushing \& McMullin. Namely, each of the particles can only be subjected to two different experiments,
and as before, each such experiment can only give one of two results. \\
In view of the specific quantum mechanical setup considered, the experiments to which the two particles are subjected
can be identified with certain angles in $[ 0, 2 \pi ]$ which define the directions along which quantum spins are
measured. As far as the results obtained, they can be identified with quantum spins, and as such will be denoted by
$+$ and $-$, respectively. Finally, when the same experiment is performed on both particles, it is assumed that due to
their entanglement and momentum conservation, the results are always different, that is, one result is $+$, while the
other is $-$. \\

Locality, as before, will mean that, when far removed in space from one another, each particle can be subjected to any
experiment independently, and the result does not depend on what happens with the other particle. \\

Having done a large number of experiments on such two particles, let us denote by

\bigskip
(4.3.1) \quad $ p_{1, 2}~ ( \alpha_i, \beta_j ~|~ x, y ) $

\medskip
the probability that experiment $\alpha_i \in [ 0, 2 \pi ]$, with $i \in \{~ 1, 2 ~\}$, on particle 1 yields the result $x
\in \{~ +, - ~\}$, and at the same time experiment $\beta_j \in [ 0, 2 \pi ]$, with $j \in \{~ 1, 2 ~\}$, on particle 2 yields the
result $y \in \{~ +, - ~\}$. \\
Similarly we denote by

\bigskip
(4.3.2) \quad $ p_1~ ( \alpha_i ~|~ x ),~~~ p_2~ ( \beta_j ~|~ y ) $

\medskip
the respective probabilities that experiment $\alpha_i \in [ 0, 2 \pi ]$, with $i \in \{~ 1, 2 ~\}$, on particle 1 yields the
result $x \in \{~ +, - ~\}$, and that experiment $\beta_j \in [ 0, 2 \pi ]$, with $j \in \{~ 1, 2 ~\}$, on particle 2 yields the result
$y \in \{~ +, - ~\}$,. \\

Now based alone on the assumption of locality, one obtains {\it Bell's inequality}

\bigskip
(4.3.3) \quad $ \begin{array}{l} - 1 ~\leq~ p_{1, 2}~ ( \alpha_1, \beta_1 ~|~ +, + ) ~+~ p_{1, 2}~ ( \alpha_1, \beta_2  ~|~ +, + ) ~+~ \\ \\
                       ~~~~~~ ~+~ p_{1, 2}~ ( \alpha_2, \beta_2 ~|~ +, +  ) ~-~ p_{1, 2}~ ( \alpha_2, \beta_1 ~|~ +, + ) ~-~ \\ \\
                       ~~~~~~ ~-~ p_1~ ( \alpha_1 ~|~ + ) ~-~ p_2~ ( \beta_2 ~|~ + ) ~\leq~ 0
                 \end{array} $

\medskip
Obviously, by changing the indices of the angles and the spin values, one can obtain further variations of this
inequality. \\

What suitable quantum mechanical experiments can give are very good approximations of the relations

\bigskip
(4.3.4) \quad $ \begin{array}{l} p_{1, 2}~ ( \alpha, \beta ~|~ +, + ) ~=~ p_{1, 2}~ ( \alpha, \beta ~|~ -, - ) ~=~
                                         ( 1 / 2 ) \sin^2 ( \alpha - \beta ) / 2 \\ \\
                                 p_{1, 2}~ ( \alpha, \beta ~|~ +, - ) ~=~ p_{1, 2}~ ( A, B ~|~ -, + ) ~=~
                                         ( 1 / 2 ) \cos^2 ( \alpha - \beta ) / 2 \\ \\
                                 p_1~ ( \alpha ~|~ + ) ~=~ p_2~ ( \beta ~|~ - ) ~=~ 1 / 2
                 \end{array} $ \\

\medskip
where $\alpha, \beta \in [ 0, \pi ]$. \\

Now let us return to the Bell inequality in (4.3.3) and take following angles for the experiments

\bigskip
(4.3.5) \quad $ \alpha_1 ~=~ \pi / 3,~~~ \alpha_2 ~=~ \pi,~~~ \beta_1 ~=~ 0,~~~ \beta_2 ~=~ 2 \pi /3 $

\medskip
in which case we obtain the {\it contradiction}

\bigskip
(4.3.6) \quad $ - 1 / 8 ~\geq~ 0 $

\medskip
As shown in Pitowsky, the Bell inequality in (4.3.3), as well as its mentioned variants follow from the Bonferroni
inequalities in (4.1.3). \\

Let us conclude the issue of Bell's inequalities, and more importantly, of the Bell Effect, by noting that the resulting
contradictions show the existence of relevant physics {\it beyond} any classical framework which obeys the principle of
locality. \\
And the quantum mechanical experiments which, together with Bell's inequalities, deliver the above contradictions are
therefore part of such a physics, even if Quantum Mechanics as a theory is still quite far from having at last settled
its foundational controversies. \\

As far as {\it entangled} quantum particles, or in general, systems are concerned, they are some of the simplest
quantum phenomena to lead to the Bell Effect, and thus beyond the classical and local framework. This is therefore one
of the reasons why they can offer possibilities in quantum computation which cannot be reached anywhere near by
usual electronic digital computers, which obviously belong to realms of physics that are classical and subjected to the
locality principle. \\ \\

{\bf 4.4 Locality versus Nonlocality}

\bigskip
The original EPR paper, then the de Broglie-Bohm causal interpretation, as well as Bell's inequalities have focused a
considerable attention on the issue of locality versus nonlocality. And in view of what appear to be obvious reasons,
there is a rather unanimous and strong dislike of nonlocality among physicists. A typical instance of such a position is
illustrated by the next citation from a letter of A Einstein to Max Born, see Maudlin, or Born :

\bigskip
\begin{quote}
... If one asks what, irrespective of quantum mechanics, is characteristic of the world of ideas of physics, one is first of
all struck by the following : the concepts of physics relate to a real outside world, that is, ideas are established relating
to things such as bodies, fields, etc., which claim "real existence" that is independent of the perceiving subject - ideas
which, on the other hand, have bee brought into as secure a relationship as possible with the sense-data. It is further
characteristic of these physical objects that they are thought of as arranged in a space-time continuum. An essential
aspect of this arrangement of things in physics is that they lay claim, at a certain time, to an existence independent of
one another, provided these objects "are situated in different parts of space". Unless one makes this kind of
assumptions about the independence of the existence (the "being-thus") of objects which are far apart from one another
in space - which stems in the first place from everyday thinking - physical thinking in the familiar sense would not be
possible. It is also hard to see any way of formulating and testing the laws of physics unless one makes a clear
distinction of this kind. This principle has been carried to extremes in the field theory by localizing the elementary
objects on which it is based and which exist independently of each other, as well as the elementary laws which have
been postulated for it, in the infinitely small (four dimensional) elements of space. \\
The following idea characterizes the relative independence of objects far apart in space (A and B) : external influence
on A has no direct influence on B; this is known as the "principle of contiguity", which is used consistently in the field
theory. If this axiom were to be completely abolished, the idea of laws which can be checked empirically in the
accepted sense, would become impossible...
\end{quote}

\medskip
However, as often happens in the case of strongly felt dislikes, the reactions involved may prove to be exaggerated.
And in the case of nonlocality this seems to happen. \\
Indeed, certain milder, fast diminishing forms of nonlocality have been around in physics, and some of them, like the
gravitational effect of a mass, were introduced by no lesser contributors than Isaac Newton. Of course, the
gravitational effect of a given mass, although it decreases fast, namely, with the square of the distance, it is
nevertheless not supposed to vanish completely anywhere. A similar thing is supposed to happen with the electric
charge, according to Culomb's law. \\
On the other hand, certain nonlocality effects in the case of entangled quantum particles are {\it not} supposed to
diminish at all with the distance separating the particles. \\

What seems to happen, however, is that there is a significant reluctance to admit even one single, and no matter how
narrow and well circumscribed instance of a {\it nondiminishing nonlocality}. Such a reluctance appears to be based on
the perception that the acceptance of even one single such nondiminishing nonlocality would instantly bring with it the
collapse of nearly all of the theoretical body of physics. \\

In other words, it is considered that physical theory, as it stands, is {\it critically unstable} with respect to the
incorporation of even one single nondiminishing nonlocality. \\

Clearly, if indeed such may be the case, then that should rather be thoroughly investigated, instead of being merely
left to perceptions as part of an attitude which, even if by default, treats it as a taboo. \\
After all, a somewhat similar phenomenon was still going on less than four centuries ago, when the idea of Galileo that
our planet Earth is moving was felt to be an instant and mortal threat to the whole edifice of established theology. \\ \\

\chapter{The Deutsch-Jozsa Algorithm}

The Deutsch-Jozsa algorithm is a good example of a quantum algorithm which by using {\it quantum parallelism} can
solve a specific problem faster than any algorithm on a usual electronic digital computer can do. To put it simply,
quantum parallelism allows certain kind of simultaneous computations, thus saving computation time. Such a feature is
not available on usual electronic digital computers, unless one sets up a special hardware with multiple circuits so that
they function in parallel and simultaneously. In the case of a quantum computer, however, certain parallel computations
are always readily available. \\

Another feature of quantum algorithms used in this chapter is {\it quantum interference}, which is not at all available on
usual electronic digital computers. \\

We shall present the Deutsch-Jozsa algorithm as a fourth step in solving certain problems, each of which leads to an
algorithm that is more involved than the previous one. \\ \\

{\bf 5.1 A simple case of quantum parallelism}

\bigskip
Suppose we are given a function $f : \{~ 0, 1 ~\} \longrightarrow \{~ 0, 1 ~\}$ which thus takes classical bits into classical
bits. Although the function $f$ has a domain of definition which only has two elements, nevertheless, its computation
can happen to be given by a complicated formula, thus it may require a large amount of computer time. Therefore, it is
convenient to avoid computing separately each of its two classical bit values $f ( 0 )$ and $f ( 1 )$, which is the only
procedure available on a usual electronic digital computer. \\

Here we shall show how by using {\it quantum parallelism} we can, through one single classical value computation,
obtain a quantum state which contains {\it both} of the classical bit values $f ( 0 )$ and $f ( 1 )$. \\

Let us start by assuming that the computation on a usual electronic digital computer of any one of the classical bit
values of the function $f$ is made by a "black box"

\begin{math}
\setlength{\unitlength}{1cm}
\thicklines
\begin{picture}(15,2)
\put(2.5,0.9){$x$}
\put(3,1){\line(1,0){1}}
\put(4,0.5){\line(0,1){1}}
\put(4,1.5){\line(1,0){1}}
\put(4,0.5){\line(1,0){1}}
\put(5,0.5){\line(0,1){1}}
\put(4.4,0.88){$\mbox{f}$}
\put(5,1){\line(1,0){1}}
\put(6.2,0.9){$f ( x )$}
\put(3.8,-0.3){$\mbox{Fig. 5.1.1}$}
\end{picture}
\end{math} \\

\medskip
Then as shown in Nielsen \& Chuang, it is possible to construct a comparably efficient quantum gate with two qubit
input and two qubit output

\begin{math}
\setlength{\unitlength}{1cm}
\thicklines
\begin{picture}(15,3)
\put(1.7,1.9){$|~ x >$}
\put(1.7,0.9){$|~ y >$}
\put(3,1){\line(1,0){1}}
\put(3,2){\line(1,0){1}}
\put(4,0.5){\line(0,1){2}}
\put(4,2.5){\line(1,0){1}}
\put(4,0.5){\line(1,0){1}}
\put(5,0.5){\line(0,1){2}}
\put(4.25,1.35){$\mbox{U}_f$}
\put(5,1){\line(1,0){1}}
\put(5,2){\line(1,0){1}}
\put(6.4,1.9){$|~ x >$}
\put(6.4,0.9){$|~ y \oplus f ( x ) >$}
\put(4,-0.3){$\mbox{Fig. 5.1.2}$}
\end{picture}
\end{math} \\

\medskip
where $x,~ y \in \{~ 0, 1 ~\}$, while as before, $\oplus$ is the addition modulo $2$. Now let us use this as a quantum
"black box" and construct with it the following quantum device

\begin{math}
\setlength{\unitlength}{1cm}
\thicklines
\begin{picture}(15,3)
\put(1.7,1.4){$|~ \psi >$}
\put(3,2){\line(1,0){1}}
\put(4,1.5){\line(0,1){1}}
\put(4,1.5){\line(1,0){1}}
\put(4,2.5){\line(1,0){1}}
\put(5,1.5){\line(0,1){1}}
\put(4.3,1.9){$\mbox{H}$}
\put(3,1){\line(1,0){3}}
\put(5,2){\line(1,0){1}}
\put(6,0.5){\line(0,1){2}}
\put(6,2.5){\line(1,0){1}}
\put(6,0.5){\line(1,0){1}}
\put(7,0.5){\line(0,1){2}}
\put(6.25,1.35){$\mbox{U}_f$}
\put(7,1){\line(1,0){1}}
\put(7,2){\line(1,0){1}}
\put(8.4,1.4){$|~ \chi >$}
\put(4,-0.3){$\mbox{Fig. 5.1.3}$}
\end{picture}
\end{math} \\

\medskip
where $H$ is the Hadamard gate in (2.2.8). \\

If we now input $|~ \psi > ~=~ |~ 0, 0 >$ then as output we obtain

\bigskip
(5.1.1) \quad $ |~ \chi > ~=~ ( 1 / \sqrt 2 )~ (~ |~ 0, f ( 0 ) > ~+~ |~ 1, f ( 1 ) > ~) $

\medskip
Here we can observe {\it quantum parallelism} in computational action. Indeed, the output state in (5.1.1) contains {\it
both} function values $f ( 0 )$ {\it and} $f ( 1 )$, although in Fig. 5.1.3 the device $U_f$ in Fig. 5.1.2, and which computes
the values of $f$, was activated only once. \\

Next we show how quantum parallelism can be used in far more powerful ways as well. \\ \\

{\bf 5.2 Massive quantum parallelism}

\bigskip
For an arbitrary integer $n \geq 1$, we shall define the $n$-fold Walsh-Hadamard quantum gate $H^{\otimes n}$ with
$n$ qubits input and $n$ qubits output as given by the $n$-fold parallel device

\begin{math}
\setlength{\unitlength}{1cm}
\thicklines
\begin{picture}(15,7)
\put(3,6){\line(1,0){1}}
\put(4,5.5){\line(0,1){1}}
\put(4,6.5){\line(1,0){1}}
\put(4,5.5){\line(1,0){1}}
\put(5,5.5){\line(0,1){1}}
\put(4.35,5.88){$\mbox{H}$}
\put(5,6){\line(1,0){1}}
\put(3,4.5){\line(1,0){1}}
\put(4,4){\line(0,1){1}}
\put(4,5){\line(1,0){1}}
\put(4,4){\line(1,0){1}}
\put(5,4){\line(0,1){1}}
\put(4.35,4.38){$\mbox{H}$}
\put(5,4.5){\line(1,0){1}}
\put(4.45,3.68){\line(0,-1){0.35}}
\put(4.45,2.98){\line(0,-1){0.35}}
\put(4.45,2.28){\line(0,-1){0.35}}
\put(3,1){\line(1,0){1}}
\put(4,0.5){\line(0,1){1}}
\put(4,1.5){\line(1,0){1}}
\put(4,0.5){\line(1,0){1}}
\put(5,0.5){\line(0,1){1}}
\put(4.35,0.88){$\mbox{H}$}
\put(5,1){\line(1,0){1}}
\put(3.8,-0.4){$\mbox{Fig. 5.2.1}$}
\end{picture}
\end{math} \\

\medskip
Clearly, $H^{\otimes n}$ is a unitary operator on the $n$-fold tensor product, see ???

\bigskip
(5.2.1) \quad $ {\bf C}^2 \otimes ~.~.~.~ \otimes {\bf C}^2 ~\simeq~ {\bf C}^{2^n} $

\medskip

And it is easy to see that if all the $n$ input qubits are $|~ 0 >$, that is, we input in Fig. 5.2.1

\bigskip
(5.2.2) \quad $ |~ 0 ~.~.~.~ 0 > ~\in {\bf C}^2 \otimes ~.~.~.~ \otimes {\bf C}^2 ~\simeq~ {\bf C}^{2^n} $

\medskip
then the $n$ qubit output will be

\bigskip
(5.2.3) \quad $ ( 1 / \sqrt 2^n )~ \Sigma_{x_1, ~.~.~.~ , x_n}~ |~ x_1, ~.~.~.~ , x_n > $

\medskip
where the sum is taken over all possible $x_1,~.~.~.~ , x_n \in \{~ 0, 1~\}$, hence it has $2^n$ terms. \\

In this way, by using only $n$ parallel Hadamard gates, the $n$-fold Walsh-Hadamard gate $H^{\otimes n}$ in Fig.
5.2.1 produces from the $n$ qubit input in (5.2.2) a {\it superposition} of no less than $2^n$ quantum states, given in
(5.2.3). \\

Let us now use this massive quantum parallelism which obviously has no correspondent in usual electronic digital
computers. \\

Given a function $f : \{~ 0, 1 ~\}^n \longrightarrow \{~ 0, 1 ~\}$ which transforms $n$ classical bits $x_1, ~.~.~.~ , x_n$ into
one classical bit $f ( x_1, ~.~.~.~ , x_n )$, we can use the above parallelism in order to evaluate this function in the
following way. Similar to the quantum gate in Fig. 5.1.2 which corresponds to the case when $n = 1$, we can construct
a quantum gate

\begin{math}
\setlength{\unitlength}{1cm}
\thicklines
\begin{picture}(15,7)
\put(1.7,5.9){$|~ x_1 >$}
\put(1.9,4.8){\line(0,-1){0.35}}
\put(1.9,4.1){\line(0,-1){0.35}}
\put(1.9,3.4){\line(0,-1){0.35}}
\put(6.6,4.8){\line(0,-1){0.35}}
\put(6.6,4.1){\line(0,-1){0.35}}
\put(6.6,3.4){\line(0,-1){0.35}}
\put(1.7,1.9){$|~ x _n >$}
\put(1.7,0.9){$|~ y >$}
\put(3,6){\line(1,0){1}}
\put(3,1){\line(1,0){1}}
\put(3,2){\line(1,0){1}}
\put(4,0.5){\line(0,1){6}}
\put(4,6.5){\line(1,0){1}}
\put(4,0.5){\line(1,0){1}}
\put(5,0.5){\line(0,1){6}}
\put(4.2,3.35){$\mbox{U}_f$}
\put(5,1){\line(1,0){1}}
\put(5,2){\line(1,0){1}}
\put(5,6){\line(1,0){1}}
\put(6.4,5.9){$|~ x_1 >$}
\put(6.4,1.9){$|~ x_n >$}
\put(6.4,0.9){$|~ y \oplus f ( x_1, ~.~.~.~ , x_n ) >$}
\put(3.8,-0.3){$\mbox{Fig. 5.2.2}$}
\end{picture}
\end{math} \\

Further, similar with the way in the particular case of $n = 1$, where we went from the device in Fig. 5.1.2 to that in Fig.
5.1.3, now we shall use the device in Fig. 5.2.2 together with the massive parallel device in Fig. 5.2.1 in order to
construct the corresponding general version of the device in Fig. 5.1.3, namely \\

\begin{math}
\setlength{\unitlength}{1cm}
\thicklines
\begin{picture}(15,3)
\put(1.2,1.4){$|~ \psi >$}
\put(2.5,2){\line(1,0){1.5}}
\put(3,1.6){\line(1,2){0.4}}
\put(3.3,2.55){$\mbox{n}$}
\put(4,1.5){\line(0,1){1}}
\put(4,1.5){\line(1,0){1.5}}
\put(4,2.5){\line(1,0){1.5}}
\put(5.5,1.5){\line(0,1){1}}
\put(4.4,1.9){$\mbox{H}^{\otimes n}$}
\put(2.5,1){\line(1,0){4.5}}
\put(5.5,2){\line(1,0){1.5}}
\put(6,1.6){\line(1,2){0.4}}
\put(6.3,2.55){$\mbox{n}$}
\put(7,0.5){\line(0,1){2}}
\put(7,2.5){\line(1,0){1}}
\put(7,0.5){\line(1,0){1}}
\put(8,0.5){\line(0,1){2}}
\put(7.25,1.35){$\mbox{U}_f$}
\put(8,1){\line(1,0){1.5}}
\put(8,2){\line(1,0){1.5}}
\put(8.5,1.6){\line(1,2){0.4}}
\put(8.8,2.55){$\mbox{n}$}
\put(9.9,1.4){$|~ \chi >$}
\put(4,-0.3){$\mbox{Fig. 5.2.3}$}
\end{picture}
\end{math} \\

\medskip
where the sign

\begin{math}
\setlength{\unitlength}{1cm}
\thicklines
\begin{picture}(15,0)
\put(3.1,0.15){\line(1,2){0.4}}
\put(3.4,1,2){$\mbox{n}$}
\put(4,0.5){$\mbox{represents}~n~\mbox{qubits.}$}
\end{picture}
\end{math}

\medskip
Now if we input in Fig. 5.2.3 the $n + 1$ qubits $|~ 0 ~.~.~.~ 0 0 >$ then we shall obtain the $n + 1$ qubit output

\bigskip
(5.2.4) \quad $ ( 1 / \sqrt 2^n )~ \Sigma_{x_1, ~.~.~.~ , x_n}~ |~ x_1, ~.~.~.~ , x_n > |~ f (  x_1, ~.~.~.~ , x_n ) > $

\medskip
which is a superposition containing {\it all} the $2^n$ different possible values of the function $f$. \\

Here it is important to note the following. This massive quantum parallelism obtained in (5.2.4) allows us to obtain {\it
simultaneously as a superposition} all the $2^n$ values of the function $f$, each of these values being a classical bit.
However, having them as a superposed quantum state given by (5.2.4), need {\it not} also mean that we can recover all
of them at once as separate classical bits. Indeed, according to the axioms of Quantum Mechanics, if we make any
measurement of the quantum state (5.2.4), and thus we obtain as value a real number, then by such a measurement we
{\it collapse} the superposed state in (5.2.4) into one, and only one, of the $2^n$ states which are the terms of the
respective sum, and we do so with the same probability $1 / \sqrt 2^n$. \\

Therefore, in order to be able to make use of the obvious immense advantages of massive quantum parallelism, we
also have to be able to find ways to {\it extract} the real numbers which are there simultaneously in superpositions,
such as for instance in (5.2.4). \\

In the next two sections we show how that can be done in the case of two specific problems. \\ \\

{\bf 5.3 The Deutsch algorithm}

\bigskip
The algorithm presented here is a modified version of the original 1985 one given by D Deutsch, see Brown, Deutsch
[1-3]. Its interest is in the fact that it uses both quantum {\it parallelism} and quantum {\it interference} in order to solve
the respective problem, and do so with a significantly better performance than a usual electronic digital computer
would do. \\
The problem is as follows. We are given, as in section 5.1, a function $f : \{~ 0, 1 ~\} \longrightarrow \{~ 0, 1 ~\}$ which
takes classical bits into classical bits. And we want to compute a {\it global} property of this function, given by the
quantity

\bigskip
(5.3.1) \quad $ f ( 0 ) \oplus f ( 1 ) $

\medskip
thus depending on {\it both} of its values, where as before $\oplus$ denotes addition modulo $2$. In other words, we
want to compute the {\it parity} of the function $f$. \\
This can be achieved with the help of the following quantum  two qubit input, two qubit output device

\begin{math}
\setlength{\unitlength}{1cm}
\thicklines
\begin{picture}(15,5)
\put(0.7,2.2){$|~ \psi >$}
\put(2,3){\line(1,0){1}}
\put(3,2.5){\line(0,1){1}}
\put(3,2.5){\line(1,0){1}}
\put(3,3.5){\line(1,0){1}}
\put(4,2.5){\line(0,1){1}}
\put(3.3,2.9){$\mbox{H}$}
\put(2,1.5){\line(1,0){1}}
\put(4,1.5){\line(1,0){1}}
\put(4,3){\line(1,0){1}}
\put(5,0.5){\line(0,1){3.5}}
\put(5,4){\line(1,0){1}}
\put(5,0.5){\line(1,0){1}}
\put(6,0.5){\line(0,1){3.5}}
\put(5.25,2.1){$\mbox{U}_f$}
\put(6,1.5){\line(1,0){3}}
\put(6,3){\line(1,0){1}}
\put(3,1){\line(0,1){1}}
\put(3,1){\line(1,0){1}}
\put(3,2){\line(1,0){1}}
\put(4,1){\line(0,1){1}}
\put(3.3,1.35){$\mbox{H}$}
\put(8,3){\line(1,0){1}}
\put(7,2.5){\line(0,1){1}}
\put(7,2.5){\line(1,0){1}}
\put(7,3.5){\line(1,0){1}}
\put(8,2.5){\line(0,1){1}}
\put(7.3,2.9){$\mbox{H}$}
\put(9.4,2.2){$|~ \chi >$}
\put(4,-0.5){$\mbox{Fig. 5.3.1}$}
\end{picture}
\end{math} \\ \\

\medskip
in which we input this time the two qubits $|~ \psi > ~=~ |~ 0, 1 >$. Here for clarity, and as in Fig. 3.2.1, let us again
decompose the above device. This time we can do so in three quantum devices, each with two qubit input and two qubit
output. The first of them is

\begin{math}
\setlength{\unitlength}{1cm}
\thicklines
\begin{picture}(15,4.5)
\put(1.7,2.2){$|~ \psi_0 >$}
\put(3,3){\line(1,0){1}}
\put(4,2.5){\line(0,1){1}}
\put(4,2.5){\line(1,0){1}}
\put(4,3.5){\line(1,0){1}}
\put(5,2.5){\line(0,1){1}}
\put(4.3,2.9){$\mbox{H}$}
\put(3,1.5){\line(1,0){1}}
\put(5,1.5){\line(1,0){1}}
\put(5,3){\line(1,0){1}}
\put(4,1){\line(0,1){1}}
\put(4,1){\line(1,0){1}}
\put(4,2){\line(1,0){1}}
\put(5,1){\line(0,1){1}}
\put(4.3,1.35){$\mbox{H}$}
\put(6.1,2.2){$|~ \psi_1 >$}
\put(4,-0.3){$\mbox{Fig. 5.3.2}$}
\end{picture}
\end{math} \\

\medskip
in which we input $|~ \psi_0 > ~=~ |~ \psi > ~=~ |~ 0, 1 >$. Then the two Hadamard gates will give

$$ |~ \psi_1 > ~=~ ( 1 / 2 )~( |~ 0 > ~+~ |~ 1 > )~( |~ 0 > ~-~ |~ 1 > ) $$

\medskip
We now input $|~ \psi_1 >$ in the following second quantum device which again has a two qubit input and a two qubit
output

\begin{math}
\setlength{\unitlength}{1cm}
\thicklines
\begin{picture}(15,5)
\put(2.7,2.2){$|~ \psi >$}
\put(3.5,3){\line(1,0){1}}
\put(3.5,1.5){\line(1,0){1}}
\put(4.5,0.5){\line(0,1){3.5}}
\put(4.5,4){\line(1,0){1}}
\put(4.5,0.5){\line(1,0){1}}
\put(5.5,0.5){\line(0,1){3.5}}
\put(4.75,2.1){$\mbox{U}_f$}
\put(5.5,1.5){\line(1,0){1}}
\put(5.5,3){\line(1,0){1}}
\put(6.5,2.2){$|~ \chi >$}
\put(4,-0.5){$\mbox{Fig. 5.3.3}$}
\end{picture}
\end{math} \\ \\

\medskip
First, let us note that in view of Fig. 5.1.2, if we input above

$$ |~ \psi > ~=~ |~ x >~ ( |~ 0 > ~-~ |~ 1 > ) / \sqrt 2 $$

\medskip
where $x \in \{~ 0, 1 ~\}$, then we obtain the two qubit output

$$ |~ \chi > ~=~ ( - 1 )^{ f ( x ) }~ |~ x >~( |~ 0 > ~-~ |~ 1 > ) / \sqrt 2 $$

\medskip
therefore, if we input $|~ \psi_1 >$, then we obtain

$$ |~ \psi_2 > ~=~  \left \{ \begin{array}{l} \pm~ ( |~ 0 > ~+~ |~ 1 > )~( |~ 0 > ~-~ |~ 1 > ) / 2~~~~  \mbox{if}~~~ f ( 0 ) = f ( 1 ) \\ \\
                                                                 \pm~ ( |~ 0 > ~-~ |~ 1 > )~( |~ 0 > ~-~ |~ 1 > ) / 2~~~~  \mbox{if}~~~ f ( 0 ) \neq f ( 1 )
                           \end{array} \right. $$

\medskip
Now we finally input $|~ \psi_2 >$ into the simple quantum device

\begin{math}
\setlength{\unitlength}{1cm}
\thicklines
\begin{picture}(15,4.5)
\put(1.7,2.2){$|~ \psi_2 >$}
\put(3,3){\line(1,0){1}}
\put(4,2.5){\line(0,1){1}}
\put(4,2.5){\line(1,0){1}}
\put(4,3.5){\line(1,0){1}}
\put(5,2.5){\line(0,1){1}}
\put(4.3,2.9){$\mbox{H}$}
\put(3,1.5){\line(1,0){3}}
\put(5,3){\line(1,0){1}}
\put(6.1,2.2){$|~ \psi_3 >$}
\put(4,0.1){$\mbox{Fig. 5.3.4}$}
\end{picture}
\end{math} \\

\medskip
we obtain the two qubit output

$$ |~ \psi_3 > ~=~  \left \{ \begin{array}{l} \pm~ |~ 0 >~( |~ 0 > ~-~ |~ 1 > ) / \sqrt 2~~~~  \mbox{if}~~~ f ( 0 ) = f ( 1 ) \\ \\
                                                                 \pm~ |~ 1 >~( |~ 0 > ~-~ |~ 1 > ) / \sqrt 2~~~~  \mbox{if}~~~ f ( 0 ) \neq f ( 1 )
                                       \end{array} \right. $$

\medskip
while in terms of Fig. 5.3.1, we have $|~ \chi > ~=~ |~ \psi_3 >$. Now we can note that

$$ f ( 0 ) ~\oplus~ f ( 1 ) ~=~  \left \{ \begin{array}{l} 0~~~~  \mbox{if}~~~ f ( 0 ) = f ( 1 ) \\ \\
                                                                              1~~~~  \mbox{if}~~~ f ( 0 ) \neq f ( 1 )
                                                      \end{array} \right. $$

\medskip
therefore, the two qubit output in Fig. 5.3.1 is

$$ |~ \chi > ~=~ \pm~ |~ f ( 0 ) ~\oplus~ f ( 1 ) >~( |~ 0 > ~-~ |~ 1 > ) / \sqrt 2 $$

\medskip
In this way, by measuring the first qubit in $|~ \chi >$, we can indeed determine the value of $f ( 0 ) ~\oplus~ f ( 1 )$, as
required in (5.3.1). \\

Let us note the following with respect to the Deutsch algorithm in Fig. 5.3.1. As far as the effect in it of {\it quantum
parallelism}, this is the same with what happened in section 5.1 in the algorithm in Fig. 5.1.3, and led to (5.1.1). \\

On the other hand, in Fig. 5.3.1 there is an additional effect which plays a role, namely, {\it quantum interference}.
Indeed, the two Hadamard gates in Fig. 5.3.2, which make up the first component in Fig. 5.3.1, give a two qubit quantum
state which in Fig. 5.3.3 leads to the two qubit output $|~ \psi_2 >$. And in giving this output, the respective input two
qubits have interfered with one another in such a way that we have now a {\it global} information on the function $f$.
And this global information has been obtained by one {\it single} activation of the component device in Fig. 5.3.3 which
computes the values of the function $f$. \\
Clearly, such an effect cannot be obtained on a usual electronic digital computer. \\

Needless to say, in the case of a massive parallelism, as for instance, in section 5.2, the possibilities for a convenient
use of quantum interference increase significantly. \\ \\

{\bf 5.4 The Deutsch-Jozsa algorithm}

\bigskip
The problem solved is as follows. Alice and Bob are again faraway from one another. Let $n \geq 1$ be a certain given
and fixed number. \\
Bob chooses any function $f : \{~ 0,~ 1,~.~.~.~ ,~ 2^n - 1 ~\} ~\longrightarrow~ \{~ 0, 1 ~\}$ which only has to satisfy the
condition that, either it is {\it constant}, or it is {\it balanced}, that is, it is equal to $0$ for half of the values in its domain,
and it is thus equal to $1$ for the other half. \\
Alice has to find out whether Bob chose a function which is constant, or on the contrary, one that is balanced. And she
has to do so with as little information exchange with Bob, as possible. \\
The only information exchange allowed between them is that Alice sends any  $x \in \{~ 0,~ 1,~.~.~.~ ,~ 2^n - 1 ~\}$ to Bob,
and Bob sends back to Alice the corresponding value $f ( x )$ of the function which he has chosen. \\

Clearly, the worse case for Alice is that she selects $2^{n - 1}$ values $x \in \{~ 0,~ 1,~.~.~.~ ,~ 2^n - 1 ~\}$, and obtains
from Bob the respective answers $f ( x )$, and all these answers have the same value. In this case Alice will have to
make one more such enquiry. Thus it may happen that Alice will need $2^n + 1$ such enquiries. \\

We can note that, each time, Alice sends Bob an information equivalent with $n$ classical bits, while each time, Bob
sends Alice one classical bit. \\
Furthermore, beyond the possible fun of the story with Alice and Bob, the above problem can correspond to a real
practical one. Indeed, let us again assume that computing a value $f ( x )$ may be very time consuming, due to the
complicated procedure which gives the function $f$. Thus, there is in such a case an important practical interest in
computing as few values of $f$ as possible, and certainly not $2^n + 1$ such values, which as seen, corresponds to the
worst case. \\

We shall show now that the Deutsch-Jozsa algorithm can always solve this problem with only {\it one} single
evaluation of the value of the function $f$. \\

This single function evaluation is of course performed through the quantum device in Fig. 5.1.2, which as mentioned,
has a comparable performance with the classical black box in Fig. 5.1.1 for the computation of values of $f$ on a
usual electronic digital computer. \\
The whole algorithm is as follows \\

\newpage

\begin{math}
\setlength{\unitlength}{1cm}
\thicklines
\begin{picture}(15,5)
\put(0,2.4){$|~ \psi >$}
\put(1,4){\line(1,0){1.5}}
\put(1.5,3.6){\line(1,2){0.4}}
\put(1.8,4.55){$\mbox{n}$}
\put(2.5,3.5){\line(0,1){1}}
\put(2.5,3.5){\line(1,0){1.5}}
\put(2.5,4.5){\line(1,0){1.5}}
\put(4,3.5){\line(0,1){1}}
\put(2.95,3.9){$\mbox{H}^{\otimes n}$}
\put(1,1){\line(1,0){1.9}}
\put(2.9,0.5){\line(1,0){1}}
\put(2.9,0.5){\line(0,1){1}}
\put(2.9,1.5){\line(1,0){1}}
\put(3.25,0.9){$\mbox{H}$}
\put(3.9,0.5){\line(0,1){1}}
\put(3.9,1){\line(1,0){1.55}}
\put(4,4){\line(1,0){1.5}}
\put(4.5,3.6){\line(1,2){0.4}}
\put(5,4.55){$\mbox{n}$}
\put(5.5,0.5){\line(0,1){4}}
\put(5.5,4.5){\line(1,0){1}}
\put(5.5,0.5){\line(1,0){1}}
\put(6.5,0.5){\line(0,1){4}}
\put(5.78,2.35){$\mbox{U}_f$}
\put(6.5,1){\line(1,0){4.5}}
\put(6.5,4){\line(1,0){1.5}}
\put(7,3.6){\line(1,2){0.4}}
\put(7.3,4.55){$\mbox{n}$}
\put(8,3.5){\line(0,1){1}}
\put(8,3.5){\line(1,0){1.5}}
\put(8,4.5){\line(1,0){1.5}}
\put(9.5,3.5){\line(0,1){1}}
\put(8.4,3.9){$\mbox{H}^{\otimes n}$}
\put(9.5,4){\line(1,0){1.5}}
\put(10,3.6){\line(1,2){0.4}}
\put(10.5,4.55){$\mbox{n}$}
\put(11,2.4){$|~ \chi >$}
\put(1.3,-0.7){$|~ \psi_0 >$}
\put(1.7,0){$\uparrow$}
\put(4.2,-0.7){$|~ \psi_1 >$}
\put(4.6,0){$\uparrow$}
\put(7.2,-0.7){$|~ \psi_2 >$}
\put(7.6,0){$\uparrow$}
\put(10,-0.7){$|~ \psi_3 >$}
\put(10.4,0){$\uparrow$}
\put(4,-2){$\mbox{Fig. 5.4.1}$}
\end{picture}
\end{math} \\ \\ \\ \\

\medskip
where we have indicated the way it is composed of three successive parts, namely, first $|~ \psi > ~=~ |~ \psi_0 >$
goes into $|~ \psi_1 >$, then it proceeds into $|~ \psi_1 >$, and at last it results in $|~ \psi_3 > ~=~ |~ \chi >$. \\
Here the respective vertical arrows $\uparrow$ are supposed to cut through the whole diagram in Fig. 5.4.1, thus
leading to three corresponding quantum gates, each with $n + 1$ input  and output qubits. \\

The $n + 1$ input qubits we use in the algorithm in Fig. 5.4.1 are given by

$$ |~ \psi > ~=~ |~ \psi_0 > ~=~ |~ 0 >^{\otimes n}~ |~ 1 > $$

\medskip
where for any quantum state $|~ \phi >$ and $n \geq 1$, we denote by $|~ \phi >^{\otimes n}$ the $n$-fold tensor product
$|~ \phi > ~\otimes~ .~.~.~ \otimes~ |~ \phi >$. \\

Now a direct computation, initiated by Alice, will give

$$ |~ \psi_1 > ~=~ \Sigma_{x_1,~.~.~.~ , x_n}~ |~ x_1,~.~.~.~ , x_n >~
                                                  ( |~ 0 > ~-~ |~ 1 > ) / \sqrt 2^{n ~+~ 1} $$

\medskip
where the sum is taken over all $x_1,~.~.~.~ , x_n \in \{~ 0, 1 ~\}$. The next step, when $|~ \psi_1 >$ goes into
$|~ \psi_2 >$, is effected by Bob, who computes his function $f$ upon the $n ~+~ 1$ input qubits $|~ \psi_1 >$, and
obtains

$$ |~ \psi_2 > ~=~ \Sigma_{x}~ ( - 1 )^{f ( x )}~ |~ x >~
                                                  ( |~ 0 > ~-~ |~ 1 > ) / \sqrt 2^{n ~+~ 1} $$

\medskip
where for brevity we denoted $x = ( x_1,~.~.~.~ , x_n )$. \\

The interesting thing to note with respect to $n ~+~ 1$ qubit state $|~ \psi_2 >$ is that it contains an information which
involves {\it all} the values of the function $f$, although it only used only {\it once} the quantum gate $U_f$ which
 computes that function. Further, the way $|~ \psi_2 >$ contains all the values of $f$ is through its {\it amplitude}
$||~ |~ \psi_2 >~ ||$. \\

Now let us see how a part of this global information on $f$ can be extracted by Alice, a part which is enough for her
to solve the problem. \\
In Fig. 5.4.1, this corresponds to going from $|~ \psi_2 >$ to $|~ \psi_3 >$, and this is done simply by having the
Walsh-Hadamard gate $H^{\otimes n}$ act on the first $n$ qubits in $|~ \psi_2 >$.






\end{document}